\documentclass[prd,showpacs,preprintnumbers,amsmath,amssymb,floatfix]{revtex4}

\usepackage{graphicx}% Include figure files

\usepackage{bm}% bold math
\usepackage{amsfonts}
\usepackage{array}
\usepackage{float}
\usepackage{epstopdf}
\usepackage{color}

\begin{document}

\title{Study on anisotropic strange stars in $f(R,T)$ gravity: An embedding approach under simplest linear functional of matter-geometry coupling}

\author{S.K. Maurya}
\email{sunil@unizwa.edu.om} \affiliation{Department of Mathematics and Physical Science,
College of Arts and Science, University of Nizwa, Nizwa, Sultanate of Oman}

\author{Abdelghani Errehymy}
\email{abdelghani.errehymy@gmail.com} \affiliation{Laboratory of High Energy Physics and Condensed Matter (LPHEMaC), Department of Physics, Faculty of Sciences A\"{i}n Chock, University of Hassan II, B.P. 5366 M\^{a}arif, Casablanca 20100, Morocco}

\author{Debabrata Deb}
\email{ddeb.rs2016@physics.iiests.ac.in} \affiliation{Department of Physics,
Indian Institute of Engineering Science and Technology, Shibpur,
Howrah 711103, West Bengal, India}

\author{ Francisco Tello-Ortiz}
\email{francisco.tello@ua.cl} \affiliation{Departamento de F\'isica,
Facultad de ciencias b\'asicas, Universidad de Antofagasta, Casilla 170,
Antofagasta, Chile}

\author{Mohammed Daoud}
\email{m$_{}$daoud@hotmail.com} \affiliation{Department of Physics, Faculty of Sciences, University of Ibn Tofail, B.P. 133, Kenitra 14000, Morocco\\
Abdus Salam International Centre for Theoretical Physics, Miramare, Trieste, Italy}

\date{\today}

\begin{abstract}
The present work is focused on the investigation of the existence of compact structures describing anisotropic matter distributions within the framework of modified gravity theories, specifically f(R,$\mathcal{T}$) gravity theory.  Additionally, we have taken f(R,$\mathcal{T}$) as a linear function of the Ricci scalar $R$ and the trace of the energy-momentum tensor $\mathcal{T}$ as  $f(R,\mathcal{T})$=$R+2\chi\mathcal{T}$,where $\chi$ is a dimensionless coupling parameter, and the Lagrangian matter $\mathcal{L}_m=-\frac{1}{3}\left(2p_{t}+p_{r}\right)$, to describe the complete set of field equations for the anisotropic matter distribution. We follow the embedding class one procedure using Eisland condition to obtain a full space-time description inside the stellar configuration.  Once the space-time geometry is specified we determine the complete solution of modified Einstein equations by using the MIT bag model equation of state $p_{r}=\frac{1}{3}\left(\rho-4B\right)$ that describes the strange quark matter (SQM) distribution inside the stellar system, where $B$ denotes a bag constant. The physical validity of our anisotropic solution is
confirmed by executing several physical tests.  It is worth mentioning that with the help of the observed mass values for the various strange star candidates we have predicted the exact radii by taking different values for $\chi$ and $B$. These predicted radii show monotonic decreasing nature as the parameter $\chi$ is moved from $-0.8$ to $0.8$ progressively. In this case, our anisotropic stellar system becomes more massive and transforms into more dense compact stars.  We also performed a detailed graphical analysis of the compact star. As a result, for $\chi < 0$, the current modified
$f(R,\mathcal{T})$ gravity seems promising to explain the observed massive compact astrophysical objects, similar to magnetars, massive pulsars, and Chandrasekhar super white dwarfs, which is not justified in the framework of general relativity.  Finally, note that when $\chi=0$ general relativity results for anisotropic matter distributions are recovered.
\end{abstract}


\maketitle

\section{Introduction}
%%%%%%%%%%%%%%%%%%%%%%%%%%%%%%%%%%%%%%
Developed by Einstein, general relativity (GR from now on) theory has been proved to be one of the most fascinating achievement of the last century.  With great observational support ~\citep{Will2005}, GR explains many issues at solar system scale, even to cosmological scales. However, numerous observational evidences from Supernova type Ia~\citep{Perlmutter1999,Bennett2003}, high redshift of supernova~\citep{Riess1998}, Planck data~\citep{Ade2014}, a large scale structure~\citep{Wood2007,Kowalski2008,Komatsu2009,Spergel2003,Tegmark2004,Abazajian2003,Abazajian2004,Abazajian2005,Hawkins2003,Verde2002}, etc., indicate accelerated expansion of our Universe. A surprising and controversial result, since GR predicts that a Universe dominated by matter or radiation accelerates negatively due to gravitational attraction. The source of this acceleration is attributed to dark energy ~\citep{Edmund2006}, a mysterious cosmic fluid, which has a uniform density distribution and a negative pressure.  The equivocal conduct of this dark energy has spurred numerous cosmologists to reveal its obvious attributes. Modified gravity theories are viewed as the good and hopeful decisions to uncover its secretive nature. Put forward by Buchdahl ~\citep{Buchdhal1970} in 1970, f(R) gravity theory was the first modified gravity theory used to explain some drawbacks introduced by Friedmann-Lema\^{\i}tre-Robertson-Walker isotropic homogeneous cosmological model, such as singularities. In the early 80's Starobinsky ~\citep{Starobinsky1980} uses this theory to address some questions in cosmic inflation models. The main ingredient of this theory is consider a family function f(R) of the $R$ Ricci scalar and allows the introduction of high order derivative terms (Lorentz invariants). So, a consequence of introducing an arbitrary function, there may be freedom to explain the accelerated expansion and structure formation of the Universe without adding unknown forms of dark energy or dark matter, modifying only the gravity sector and not the matter one. Nowadays, different procedures have been suggested so as to modify Einstein gravity in this ways. Qadir et al.~\citep{Qadir2017} examined different parts of adjusted relativistic elements and suggested that general relativity may needs to change in order to determine different cosmological issues, similar to quantum gravity and dark matter issue. Although in the earliest times, an obscure gravitational hypothesis described the development of the universe, it is widely recognized that adjusted gravity, which is a classical generalization of general relativity, can clarify early-time expansion and late time accelerating without presenting any type of dark elements.

The modified gravity theories are the global models appeared by changing just the gravitational part of the Einstein-Hilbert action, for further surveys on modified gravity and dark energy, see, for example,~\citep{Capozziello2010,Capozziello2011,Bamba2012,Koyama2016,Bamba2013,Yousaf2016a,Yousaf2016b,Nojiri2006,Nojiri2007,Nojiri2008a,Nojiri2008b,Sotiriou2010}. Nojiri and Odintsov~\citep{Nojiri2003} proposed the primary theoretical and observational feasible hypothesis of our quickening cosmos from $f(R)$ gravity. There has been a fascinating discussion on the formation of structures and the elements of different heavenly bodies in a ${\Lambda}$-dominated era~\citep{Yousaf2017}, $f(R)$~\citep{Sharif2015}, $f(R,\mathcal{T})$ gravity~\citep{harko2011}, where $\mathcal{T}$ is the trace of the stress energy tensor. Lately, Nojiri et al.~\citep{Nojiri2017} have investigated an assortment of enormous issues, as skipping cosmology, early-time, late-time enormous acceleration. They accentuated that some all-encompassing gravity hypotheses, for instance, $f(\mathcal{G})$, where $\mathcal{G}$ is the Gauss-Bonnet parameter, $f(R)$, where $R$ is the Ricci scalar and $f(\mathbb{T})$, where $\mathbb{T}$ is the torsion scalar, can be demonstrated to disclose different fascinating enormous scenarios.

The search for the impacts of anisotropies on the distributions of matter of astrophysical objects is the fundamental key that incites the different fascinating phenomena, similar to presence of a strong just as Minkowskian center~\citep{Herrera2008,DiPrisco2011}, buildup of pions~\citep{Sawyer1972}, changes of phase of various kinds~\citep{Sokolov1980} etc. One can compose all conceivable precise solutions of the static relativistic crumbling cylinder in the isotropic nature in terms of standard expressions in general relativity~\citep{Herrera2008} as well as in $f(R)$ gravity~\citep{Sharif2015}. Sussman and Jaime~\citep{Sussman2017} have decomposed a class of irregular spherical arrangements in view of a particular trace-free anisotropic pressure tensor for the decision of $f(R)=R^{1/2}$ gravity model. Shabani and Ziaie~\citep{Shabani2017} utilized numerical and dynamical procedures to examine the impacts of a specific $f(R,\mathcal{T})$ gravity model on the steadiness of the emerging Einstein universe. Garattini and Mandanici~\citep{Garattini2017} analyzed some steady structures of different anisotropic relativistic astrophysical objects and inferred that additional curvature gravitational terms originating from rainbow's gravity prone to help different configurations of astrophysical stars. Sahoo et al.~\citep{Sahoo2017} investigated different cosmological viewpoints in the setting anisotropic relativistic foundations.

Recently Harko and his collaborators~\citep{harko2011} exhibited an increasingly general type of $f(R)$ gravity theory by selecting the matter Lagrangian comprises of a self-assertive expression of the Ricci scalar $R$ and the trace of the stress energy tensor given as $f(R,\mathcal{T})$. This is called the $f(R,\mathcal{T})$ gravity theory. The idea of Harko et al.~\citep{harko2011} consisted in verifying whether new materials, rather than geometrically specific terms, were able to escape the gravity-related problems of $f(R)$ when they introduced $f(R,\mathcal{T})$ gravity. This new $f(R,\mathcal{T})$ gravity theory has been effectively linked to describe the late time acceleration of the expansion of the universe~\citep{Shabani2014,Baffou2015,Moraes2017} as well as to explore various problems related to stellar astrophysics~\citep{Sharif2014,Alhamzawi2015}.

Then again, Wu et al.~\citep{Wu2018} and Barrientos et al.~\citep{Barrientos2018} introduced the Palatini way to deal with the detailing of $f(R,\mathcal{T})$ theory of gravity by enabling the association to be free of metric. Further, the global shape of the hydrostatic equilibrium equation named as Tolman-Oppenheimer-Volkoff (TOV) equation for the anisotropic and isotropic astrophysical objects, in the frame of the $f(R,\mathcal{T})$ theory of gravity was introduced in the following literatures~\citep{Moraes2016,Deb2018}. The investigations of Moraes et al.~\citep{Moraes2017} and Das et al.~\citep{Das2017} have discovered the effect of $f(R,\mathcal{T})$ gravity theory on astrophysical configurations, such as wormholes and gravitational vacuum stars, respectively.

Harko et al.~\citep{harko2011} in their spearheading work referenced that the inspiration driving considering T-reliance in the $f(R,\mathcal{T})$ gravity theory is the conceivable presence of exotic non-ideal fluids or impacts of quantum, for instance, the particle construction~\citep{Harko2014}. The authors in their investigation~\citep{harko2011} demonstrated that the covariant derivative of the stress-energy tensor has not disappeared and that additional acceleration will be reliably available in $f(R,\mathcal{T})$ gravity because of the coupling between terms of the curvature and the matter. Therefore, in the context of the $f(R,\mathcal{T})$ gravity theory the particles was followed a non-geodesic path. Subsequently, this question was addressed by Chakraborty~\citep{SC2013} and demonstrated that for a specific function of $f(R,\mathcal{T})$, like $f(R,\mathcal{T})=R+h(\mathcal{T})$, the particles tested follow the geodesic path. Consequently, the author~\citep{SC2013} exhibited that the entire structure would act like a non-interacting two-fluid structure where the second sort of fluid is produced because of the interaction between the matter and the geometry. The main objective of this study~\citep{SC2013} is to conserve the effective stress-energy tensor with the geodesic movement of particles, within the framework of the $f(R,\mathcal{T})$ theory of gravity. Shabani and Farhoudi~\citep{Shabani2014} also examined that the $f(R,\mathcal{T})$ theory of gravity was quickly subjected to the solar system test. Similarly, it did reveal the deviation from the gravitational lens test~\citep{Baffou2017} and the typical geodetic equation~\citep{Alhamzawi2015}. A last investigation by Zaregonbadi and his colleague~\citep{Zaregonbadi2016} reveals that the $f(R,\mathcal{T})$ theory of gravity can appropriately clarify the galactic impacts of the dark matter.

Ruderman's initial work~\citep{Ruderman1972} showed that the high density of nuclear matters that interacts in a relativistic way is the main reason for the formation of anisotropy. As indicated by Ruderman, the internal pressure of the deeply compact stellar objects can be decomposed into two sections: the radial pressure $p_{r}$ and the tangential pressure $p_{t}$, where $p_{t}$ is perpendicular to $p_{r}$, i.e., anisotropy, similar an X-ray buster 4U 1820-30, PSR J1614 2230, X-ray pulsar, LMC X-4, the millisecond pulsar SAX J1804.4-36580, Her X-1, etc.. We would like to emphasize the fact that, in the context of general relativity, we can find a large number of works~\citep{Ivanov2002,Maurya2016,Maurya2017,Rahaman2010,Rahaman2011,Errehymy2019,Usov2004,MH2003,Varela2010,SM2003,Deb2016a,Rahaman2012,Shee2016} in which the impact of anisotropy on compact objects with static spherical symmetry has been examined. It should be noted that when the radial part of the pressure, $p_{r}$, varies from the angular part, $p_{\theta}=p_{\Phi}=p_{t}$, it can be said that the frame is of an anisotropic nature. It is clear that the condition $p_{\theta}=p_{\Phi}$ increases due to the impact of spherical symmetry. In a physical frame, when the scalar field associated with a spatial gradient is different from zero, the pressures are anisotropic.

Ultra-dense compact stellar stars consist of quark matter $(u)$, $(d)$ and $(s)$ are called strange stars. Witten~\citep{Witten1984} and Bodmer~\citep{Bodmer1971} conjectured the conceivable presence of strange stars with the strange quark matter as the outright fundamental state of matter in strong interaction. This new kind of compact stellar stars has attracted consideration of numerous scholars~\citep{Baym1976,Haensel1986,Alcock1986,Drago1996}. The circumstance radically modified after the collection of many of observational information was gathered by utilizing the new production $\gamma$ and X-ray satellites. It was demonstrated that specific object applicants as the conceivable candidates for strange stars are Her X-1 and 4U 1820-30~\citep{Li1995,Bombaci1997,Dey1998}. Gangopadhyay and his collaborators~\citep{Gangopadhyay2013} examined twelve conceivable candidates of spherical strange objects and anticipated their radius by utilizing the mass of compact astrophysical stars observed. Demorest et al.~\citep{Demorest2010} decided the mass of the strange star candidate PSR J1614+2230 by utilizing Shapiro delay and anticipated that such the high mass astrophysical items can only be supported by utilizing MIT bag model equation of state.

In this work, we use the framework of embedding class I techniques to embedded a 4-dimensional space-time into a 5-dimensional flat Euclidean space, so as to obtain a solution to the modified Einstein field equations in the framework of $f(R,\mathcal{T})$ gravity theory. So, this paper is organized as follows, Sec. II is devoted to the mathematical framework of our $f(R,\mathcal{T})$ gravity theory. In Sec. III, we express the basic Einstein field equations for anisotropic matter distributions in $f(R,\mathcal{T})$ gravity and present the general solutions of anisotropic strange star of class I spacetime in Sec. IV. In Sec. V, we analyze all the necessary prerequisites that an anisotropic solution of the Einstein field equations in the framework $f(R,\mathcal{T})$ theory gravity must meet to be physically allowable. Finally, in SEC. VI conclusions are reported.

\section{Basic mathematical formulation of $f(R,\mathcal{T})$ Theory}\label{sec1}

Let us assume that the action for the modified theories of gravity $f(R, \mathcal{T})$ of the following form~\citep{harko2011}

 \begin{equation}\label{1.1}
S=\frac{1}{16\pi}\int f(R,\mathcal{T})\sqrt{-g}\, d^{4}x+\int \mathcal{L}_m\sqrt{-g}\,d^{4}x,
\end{equation}

where $f(R,\mathcal{T})$ is an arbitrary function of the Ricci scalar $R$
and the trace $\mathcal{T}$ of the energy-momentum tensor $T_{\mu\,\nu}$. The tensor density $g$ is the determinant of the metric tensor $g_{\mu\nu}$ while $\mathcal{L}_m$ is Lagrangian density of the matter.
In our study we chose the geometrical units $G=c=1$.

By Varying the action $S$ of the gravitational field with respect to the metric tensor $g_{\mu\nu}$ yields the following field equation
\begin{eqnarray}\label{1.2}
&\qquad\hspace{-1cm}  \left( R_{\mu\nu}- \nabla_{\mu} \nabla_{\nu} \right)\,f_R (R,\mathcal{T}) +\Box \,f_R (R,\mathcal{T})\,g_{\mu\nu} - \frac{1}{2} f(R,\mathcal{T})\,g_{\mu\nu}   = 8\pi T_{\mu\nu} - f_\mathcal{T}(R,\mathcal{T}) \left(T_{\mu\nu}  +\Theta_{\mu\nu}\right),\end{eqnarray}
where $f_R (R,\mathcal{T})$ denote the partial derivative of $f (R,\mathcal{T})$ with respect to $R$ while $f_\mathcal{T} (R,\mathcal{T})$ is the partial derivative of $f (R,\mathcal{T})$ with respect to $\mathcal{T}$. $R_{\mu\nu}$ is the Ricci tensor,  ${\Box \equiv\partial_{\mu}(\sqrt{-g} g^{\mu\nu} \partial_{\nu})/\sqrt{-g}}$ is the D'Alambert operator,
 and $\nabla_\mu$ represents the
covariant derivative which is associated with the Levi-Civita connection of metric tensor $g_{\mu\nu}$. It should be noted that in the case where $T_{\mu\nu}$ describes a perfect fluid, the presence of terms such as $\nabla_{\mu}\nabla_{\nu}R$ and $(\nabla_{\mu}R)(\nabla_{\nu}R)$ and those that come from sector $\mathcal{T}$, prevent $T_{\mu\nu}$ from following the perfect fluid behavior. In the case of  $f(R)$ gravity ~\citep{Capozziello2019} applied to the cosmological context, to correct such deviation it is necessary to impose certain conditions on $\nabla_{\mu}\nabla_{\nu}R$ and $(\nabla_{\mu}R)(\nabla_{\nu}R)$. In the present case these conditions must also extend to the contributions of $\mathcal{T}$ to the field equations. Although the present model considers an  anisotropic matter distribution, these considerations must be addressed in order to obtain a stellar or cosmological model congruent with the given matter distribution.

So, the stress energy tensors $T_{\mu\nu}$ and $\Theta_{\mu\nu}$ are defined as follows~\citep{Landau2002},
\begin{eqnarray}
T_{\mu\nu}=g_{\mu\nu}\mathcal{L}_m-2\partial\mathcal{L}_m/\partial g^{\mu\nu}\\
\Theta_{\mu\nu}= g^{\alpha\beta}\delta T_{\alpha\beta}/\delta g^{\mu\nu}
\end{eqnarray}

The equation for covariant divergence of the stress-energy tensor $T_{\mu\nu}$ is obtained from Eq. (\ref{1.2}) as ~\citep{barrientos2014}

\begin{eqnarray}\label{1.3}
&\qquad\hspace{-2cm}\nabla^{\mu}T_{\mu\nu}=\frac{f_\mathcal{T}(R,\mathcal{T})}{8\pi -f_\mathcal{T}(R,\mathcal{T})}[(T_{\mu\nu}+\Theta_{\mu\nu})\nabla^{\mu}\ln f_\mathcal{T}(R,\mathcal{T})  +\nabla^{\mu}\Theta_{\mu\nu}-(1/2)g_{\mu\nu}\nabla^{\mu}\mathcal{T}],
\end{eqnarray}
where Eq. (\ref{1.3}) shows that the stress-energy tensor $T_{\mu\nu}$ in $f(R,T)$ gravity is not conserved, as in other theories of gravity \citep{zhao/2012,yu/2018}.

As note that the dependence on $\mathcal{T}$ can be persuaded by exotic imperfect fluids or quantum effects (conformal anomaly). In the present study we are choosing  the energy-momentum tensor for the anisotropic matter distribution of the form,
\begin{equation}\label{1.4}
T_{\mu\nu}=(\rho+{p_t})u_\mu u_\nu-{p_t}g_{\mu\nu}+\left({p_r}-{p_t}\right)v_\mu v_\nu,
\end{equation}
where ${v_{\mu}}$ is the radial-four vectors while ${u_{\nu}}$ is four velocity vectors. Here, $\rho$ is matter density, $p_r$ and $p_t$ represent the radial and tangential pressures, respectively. We have considered the Lagrangian matter $\mathcal{L}_m=-\mathcal{P}$ and  $\mathcal{P}=\frac{1}{3}(2\,p_t+p_r)$ throughout in our study. The energy tensor  $\Theta_{\mu\nu}$ for anisotropic fluid is defined as $\Theta_{\mu\nu}=-2T_{\mu\nu}-\mathcal{P} g_{\mu\nu}$.

In the present study we have assumed the functional  form of $f(R,\mathcal{T})$ as $f(R,\mathcal{T})=R+2\,f(\mathcal{T})$ ~\cite{harko2011}, where $f(\mathcal{T})$ is an arbitrary function of the trace $(\mathcal{T})$ of the stress-energy tensor of matter. In order to determine the stress-energy momentum tensor for modified theory of gravity  we choose $f(\mathcal{T})=\chi\,\mathcal{T}$, where $\chi$ is a constant.

Then linear functional form of $f(R,\mathcal{T})$ can be written as
\begin{eqnarray}
f(R,\mathcal{T})=R+2\,\chi\,\mathcal{T} \label{frt}
\end{eqnarray}

The above linear functional has been used successfully in other different $f(R,\mathcal{T})$ gravity models~\citep{singh2015,moraes2014b,moraes2015a,moraes2016b,moraes2017,reddy2013b,kumar2015,shamir2015,Fayaz2016}. By plugging the value $f(R,\mathcal{T})$ from Eq.(\ref{frt}) in Eq.~(\ref{1.2}) we obtain
\begin{eqnarray}\label{1.5}
G_{\mu\nu}=8\pi T_{\mu\nu}+\chi \mathcal{T}g_{\mu\nu}+2\chi(T_{\mu\nu}+\mathcal{P} g_{\mu\nu})=8\pi T^{eff}_{\mu\nu}
\end{eqnarray}
 where $G_{\mu\nu}$ is the usual Einstein tensor and $T^{eff}_{\mu\nu}$ is the effective energy-momentum tensor for modified theory of gravity. we note that the Einstein tensor for modified theory of gravity reproduce the standard GR results if $\chi=0$. Now the effective energy-momentum tensor $T^{eff}_{\mu\nu}$ in $f(R,\mathcal{T})$ gravity can be expressed as

   \begin{eqnarray}\label{1.5a}
   T^{eff}_{\mu\nu}= T_{\mu\nu} \left(1+\frac{\chi}{4\,\pi}\right)+\frac{\chi}{8\pi}\,(\mathcal{T}+2\,\mathcal{P})g_{\mu\nu}
   \end{eqnarray}

By inserting the value of $f(R,\mathcal{T})=R+2\chi\mathcal{T}$ in Eq.~(\ref{1.3}) we get
 \begin{eqnarray}\label{1.6}
 \nabla^{\mu}T_{\mu\nu}=-\frac{1}{2\,\left(4\pi+\chi\right)}\chi\left[g_{\mu\nu}\nabla^{\mu}\mathcal{T}+2\,\nabla^{\mu}(\mathcal{P} g_{\mu\nu})\right].
 \end{eqnarray}
Then from Eq.(\ref{1.6}), we can write the conservation of effective energy momentum tensor $T^{eff}_{\mu\nu}$ as,
\begin{eqnarray}
 \nabla^{\mu} T^{eff}_{\mu\nu} =0
\end{eqnarray}

\section{Einstein's field equations for anisotropic matter in $f\left(R,\mathcal{T}\right)$ gravity}\label{sec2}
We assume the interior spacetime of static stellar configuration is`spherically symmetric which can be described by the following line element
 \begin{equation}\label{2.1}
ds^2=e^{\nu(r)}dt^2-e^{\lambda(r)}dr^2-r^2(d\theta^2+\sin^2\theta d\phi^2),
\end{equation}
 where $\nu$ and $\lambda$ are the gravitational metric potentials which are the functions of the radial coordinate, $r$ only.

 Now using Eqs.~(\ref{1.4}), (\ref{1.5}) and (\ref{2.1}) the explicit form of Einstein field equations for $f\left(R,\mathcal{T}\right)$ gravity can be given as
  {\small{\begin{eqnarray} \label{2.3}
&\qquad\hspace{-1.2cm} {{\rm e}^{-\lambda}} \left( {\frac {\nu^{{\prime}}}{r}}+\frac{1}{{r}^{2}} \right) -\frac{1}{{r}^{2}}=8\,\pi \,p_{{r}}+\frac{\chi}{3}\, \left(- 3\,\rho+7\,p_{{r}}+2\,p_{{t}}
 \right) =8\pi{p^{{\it eff}}_r},\\ \label{2.4}
&\qquad\hspace{-3cm} \frac{{{\rm e}^{-\lambda}}}{2} \left( \nu^{{\prime\prime}}+\frac{1}{2}\,{{\nu}^{{\prime}}}^{2}+{\frac {\nu^{{\prime}}-\lambda^{{\prime}}}{r}}-\frac{1}{2}\,\nu^{{\prime}}\lambda^{{\prime}} \right) =8\,\pi \,p_{{t}}+\frac{\chi}{3}\, \left( -3\,\rho+p_{{r}}+8\,p_{{t}} \right) =8\pi{p^{{\it eff}}_t}\\
\label{2.2}
 &\qquad\hspace{-1cm} {{\rm e}^{-\lambda}} \left( {\frac {\lambda^{{\prime}}}{r}}-\frac{1}{{r}^{2}}
 \right) +\frac{1}{{r}^{2}}= 8\,\pi \,\rho+\frac{\chi}{3}\,\left( 9\,\rho-p_{{r}}-2\,p_{{t}} \right) =8\pi{\rho}^{{\it eff}}.
 \end{eqnarray}}}
where the prime $(\prime)$ denotes the differentiation with respect to the radial coordinate. Here, ${\rho}^{{\it eff}}$,~${p^{{\it eff}}_r}$~and~${p^{{\it eff}}_t}$ represent corresponding to the effective density, effective radial pressure and effective tangential pressure of the compact stellar system respectively, which are as follows:
\begin{eqnarray}\label{2.4a}
8\,\pi\,p_{{r}}-{\frac {\chi\,}{3}}\left( 3\,\rho-7\,p_{{r}}-2\,p_{{t}} \right)&=&8\,\pi\,{{p}^{{\it eff}}_r},\\\label{2.3a}
8\,\pi\,p_{{t}}-{\frac {\chi\,}{3 }}\left( 3\,\rho-p_{{r}}-8\,p_{{t}} \right) &=&8\,\pi\,{{p}^{{\it eff}}_t},\\
8\,\pi\,\rho+{\frac {\chi\,}{3}}\left( 9\,\rho-p_{{r}}-2\,p_{{t}} \right)&=&8\,\pi\,{{\rho}^{{\it eff}}}\label{2.2a}
\end{eqnarray}

Since the Eqs.~(\ref{2.3})-(\ref{2.2}) having five unknowns, Therefore in order to solve the following equations we assume that the equation of state (EOS) inside the stellar system to be governed by the well known MIT bag EOS~\citep{Chodos1974}.

 In the present study, we expect that the strange quark matter distribution is administered by the equation of state of MIT bag model. For simplicity, it is to be viewed as that $(u)$, $(d)$ and $(s)$ quarks also does not have a mass or interact in their nature. Consequently, as indicated by the MIT bag model, the quark pressure ${p_r}$ can be written as
 \begin{equation}
{p_r}={\sum_{f=u,d,s}}{p^f}-{B}, \label{eqeos1}
\end{equation}
where ${p^f}$
depects the pressure exerted on the $(u)$, $(d)$ and $(s)$ quark flavors and $B$ corresponds to the energy of the vaccum density of each flavor, called "bag consant". It is given that the energy density ${\rho}^f$ of the individual quark flavor is related to individual pressure $p^f$ by means of the relation
 \begin{equation}
p^f=\frac{1}{3}{{\rho}^f}. \label{eqeos2}
\end{equation}

On the other hand, the energy density due to the strange quarks matter distribution in the context of the bag model is given by
 \begin{equation}
{{\rho}}={\sum_{f=u,d,s}}{{\rho}^f}+B. \label{eqeos3}
\end{equation}

Presently, by utilizing Eqs.~(\ref{eqeos2}) and (\ref{eqeos3}) into Eq.~(\ref{eqeos1}) one may infer the notable equation of state of MIT bag model for strange quark matter given as
 \begin{eqnarray}\label{2.5}
 p_r=\frac{1}{3}\left(\rho-4B\right).
 \end{eqnarray}

To show the numerical result of the current model we performed all the estimates taking into account the values of $B$ as $64~MeV/{{fm}^3}$ and $74~MeV/{{fm}^3}$~\citep{Rahaman2014}. It merits referencing that these values of $B$ in our studies fall within the range of conceivable values $B$ as presented by different authors~\citep{Alcock1986,Burgio2002,Farhi1984}.

The mass function of the spherically symmetric astrophysical system is given by
\begin{equation}\label{2.6}
m \left( r \right) =4\,\pi\int_{0}^{r}\!{{\rho}^{\it eff}} \left( r \right) {r}^{2}{dr}.
\end{equation}

The combination of Eq.~(\ref{2.6}) into Eq.~(\ref{2.2}) leads to
\begin{eqnarray}\label{2.8}
 {{\rm e}^{-\lambda \left( r \right) }}=1-{\frac {2m}{r}},
 \end{eqnarray}
where $m$ is the gravitational mass inside the radius of the sphere.

\section{The general solutions of anistropic strange star of class one spacetime in $f(R,\mathcal{T})$ gravity} \label{sec3}

 \subsection{Basic formulation of class one space-time}

~\citet{Eisenhart1925} shows that a space $V^{n+1}$ can represent an embedding class 1 space i.e.  $(n+1)$ dimensional space $V^{n+1}$  can be embedded into a pseudo-Euclidean space $E^{n+2}$ of dimension $(n+2)$, if there exists a symmetric tensor $b_{mn}$ which satisfies the following  Gauss- Codazzi equations: \\
\begin{eqnarray}\label{eqcls1.1}
R_{mnpq}=2\,e\,{b_{m\,[p}}{b_{q]n}}~~~ \text{and}~~~b_{m\left[n;p\right]}-{\Gamma}^q_{\left[n\,p\right]}\,b_{mq}+{{\Gamma}^q_{m}}\,{}_{[n}\,b_{p]q}=0,
\end{eqnarray}

where $e=\pm1$, $R_{mnpq}$ denotes curvature tensor and square brackets represent antisymmetrization. Here, $b_{mn}$ are the coefficients of the second differential form. From Eq.~\ref{eqcls1.1}, \citet{Eiesland1925} derived a necessary and sufficient condition for the embedding class I in a more simplest form as
 \begin{eqnarray}
R_{{0101}}R_{{2323}}=R_{{0202}}R_{{1313}}-R_{{1202}}R_{{1303}}.\label{3.2}
\end{eqnarray}

The components of Riemannian symbols for the  spherically symmetric interior spacetime (\ref{2.1}) are given as

$ R_{{0101}}=-\frac{1}{4}\,{{\rm e}^{\nu}} \left( -\nu^{{\prime}}\lambda^{{\prime}}+{\nu^{{\prime}}}^{2}+2
\,\nu^{{\prime\prime}} \right),~~R_{{2323}}=-{r}^{2} {\sin^2 \theta} \left( 1-{{\rm e}^{-\lambda}} \right),~~
  R_{{0202}}=-\frac{1}{2}\,r\nu^{{\prime}}{{\rm e}^{\nu-\lambda}},\\ ~~ R_{{1313}}=-\frac{1}{2}\,\lambda^{{\prime}}r \sin^2 \theta,~~ R_{{1202}}=0,~~~ R_{{1303}}=0$.\\

Substituting the above components of Riemannian symbols into Eq.~(\ref{3.2}) we get the a differential equation of the form
\begin{eqnarray}\label{3.3}
({\lambda}^{{\prime}}-{{\nu}^{{\prime}
}})\,{\nu}^{{\prime}}\,{{\rm e}^{\lambda}}+2\,(1-{{\rm e}^{\lambda}}){\nu}^{{\prime\prime}}+{{\nu}^{{\prime}}}^{2}=0.
\end{eqnarray}

\subsection{General solution for anisotropic strange star}

As we can see that the field Eqs.~(\ref{2.2})-(\ref{2.4}) depends on metric function $\lambda(r)$ and $\nu(r)$. For this purpose we assume ansatz of the metric potential ${\nu}$ of the form
 \begin{equation}\label{3.1}
 {\nu(r)}=2\,A\,r^2 +\ln\,C,
 \end{equation}
 where  $A$ and $C$ are positive constants, and dimension of $A$ is $length^{-2}$. Here $e^{\nu}=\ln\,C$ at $r \rightarrow 0$, which shows that the metric potential we choose in  Eq.(\ref{3.1}) is regular and finite at centre.
On the other hand, Lake \cite{Lake2003} has performed that for any physically valid solution the gravitational potential ${{\nu}}$ must be positive and monotonically increasing function of the radial coordinate and be regular throughout within the stellar model. The gravitational metric potential we choose in Eq.~(\ref{3.1}) is satisfying all the above requirements which leads the primary physical acceptability of solution. Now in order to determine the function $\lambda(r)$ we use embedding class one condition (\ref{3.3}). For this purpose, we put the value of $e^{\lambda}$ from Eq.~(\ref{3.1}) into Eq.~(\ref{3.3}) and we obtain $\lambda$ as
 \begin{equation}\label{3.4}
 {\lambda(r)}=\ln \left[1+B\,{v^{\prime}}^2\,e^{\nu}\right]=\ln \left[1+D\,Ar^2\,e^{2\,Ar^2}\right],
 \end{equation}
 where $D=16\,C\,A\,B$ and $C$ is the arbitrary integration constant. Moreover, the following form of metric function (\ref{3.1}) which has been utilized by Maurya et al. \cite {Maurya1} to construct the well behaved relativistic charged compact star models.

\begin{figure}[h!]
\begin{center}
\includegraphics[width=7cm]{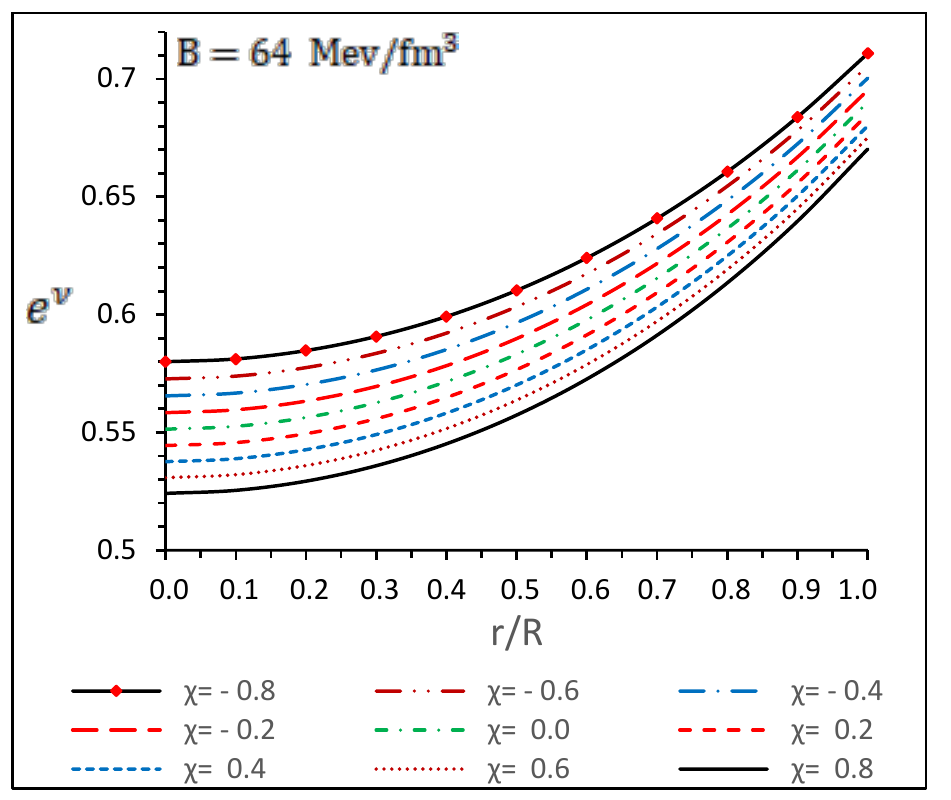}
\includegraphics[width=7cm]{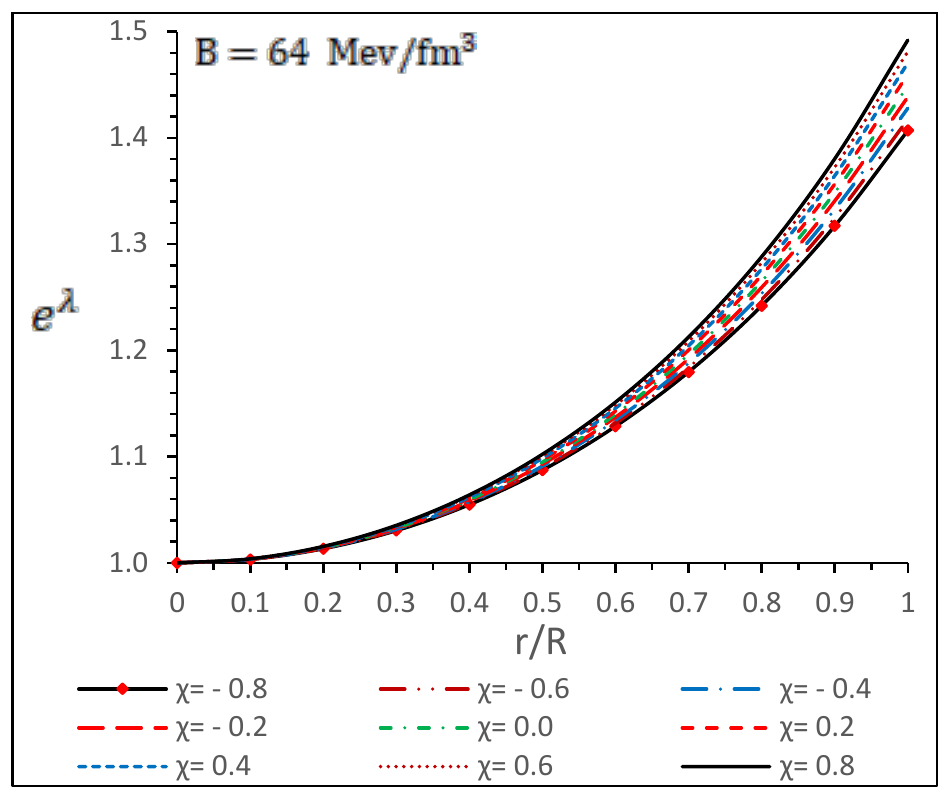}
\includegraphics[width=7cm]{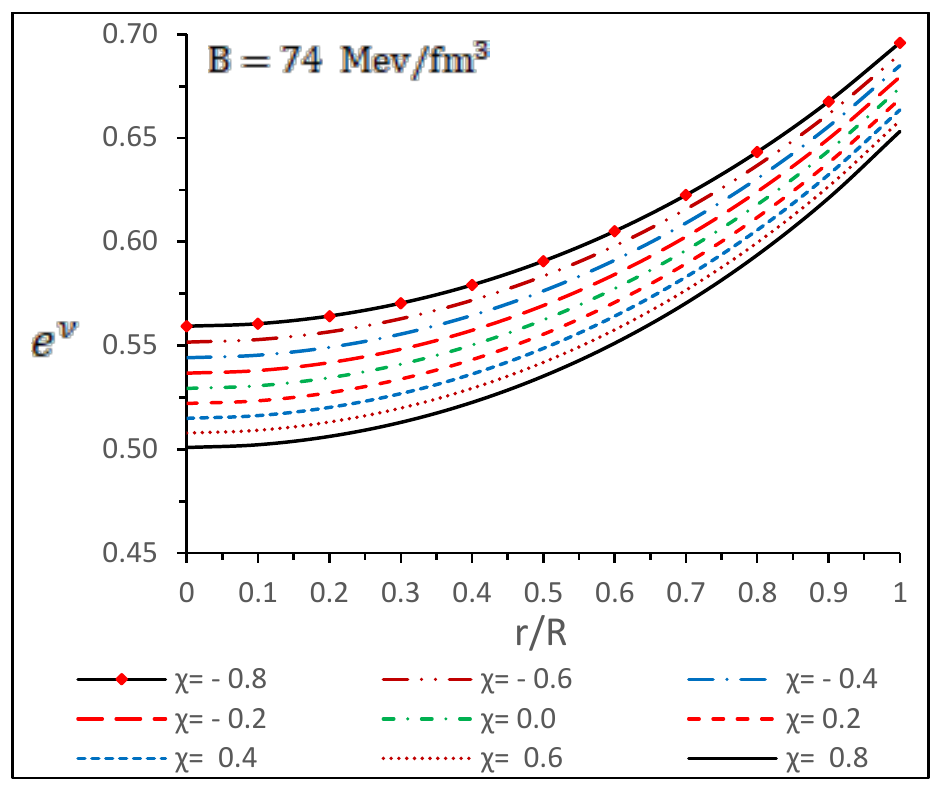}
\includegraphics[width=7cm]{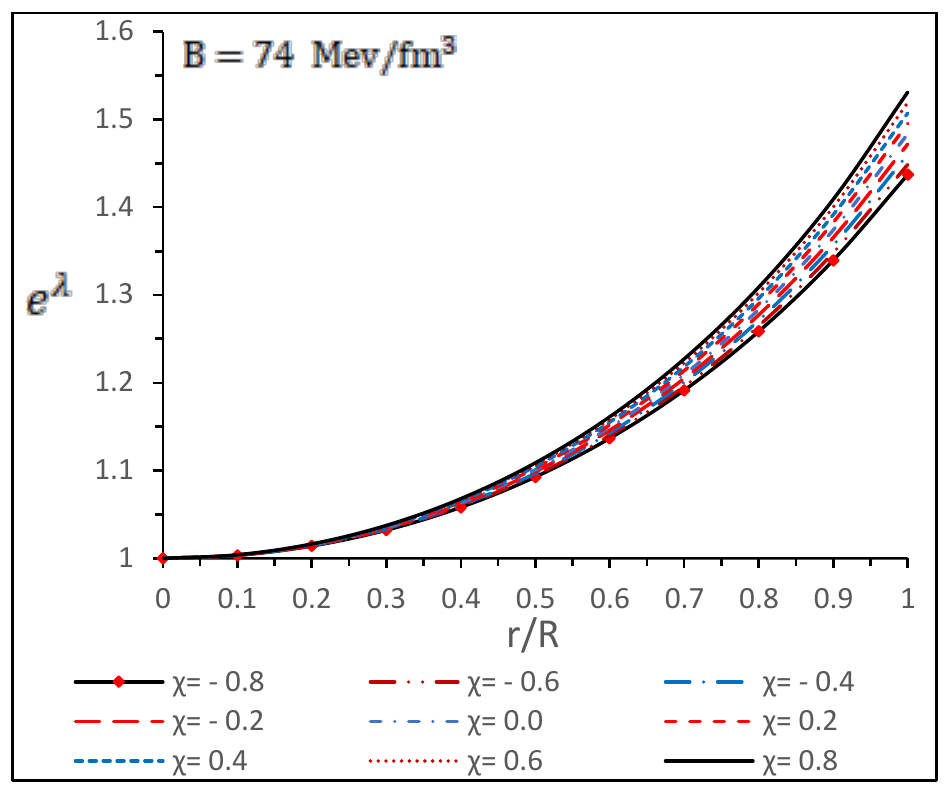}
\caption{Behavior of metric potentials vs. radial coordinates $r/R$ of SMC X-4 for different values of $\chi$ with bag constant $B=64 Mev/fm^3$ and $B=74 Mev/fm^3$.}.
\label{Fig1}
\end{center}
\end{figure}

The behavior of the metric potentials, viz., ${\rm e}^{\nu}$ and ${\rm e}^{\lambda}$, with respect to the radial coordinate is shown in Fig.~\ref{Fig1}, which shows that the metric functions are free from geometrical singularity.
%%%%%%%%%%%%%%%%%%%%%%%%%%%%%%%%%%
%%%%%%%%%%%%%%%%%%%%%%%%%%%%%%%%%%%%%%%
\subsection{Matching conditions at boundary}\label{sec4}
In order to obtain the complete set of constant parameters that characterize the model $\{A,B,C,D\}$ and the macro physical observables \i.e the mass $M$ and the radius $R$ of the anisotropic relativistic fluid sphere, one needs to match the inner geometry $\mathcal{M}^{-}$ with the corresponding exterior space-time $\mathcal{M}^{+}$ in a smoothly way. However, in the context of modified gravity theories this is a highly non-trivial issue. In the framework of general relativity theory concerning the study of compact structures the exterior manifold corresponds to (according to the case) i) Schwarzschild vacuum solution, ii) Reissner-Nordstrom iii) Kerr-Newman, etc. to name few. In the present model we are dealing with uncharged anisotropic matter distribution, it means that in principle we can join the internal geometry given by (28)-(29) with exterior Schwarzschild space-time. Nevertheless, the contributions coming from the geometrical and matter sector introduced by the $f(R,\mathcal{T})$ model can modify the external space-time surrounding the stellar interior, even introduce exotic behaviors on the surface that prevent an adequate joint. Moreover, vacuum solutions in the scenario of modified gravity theories do not necessarily coincide with those of general relativity. On the other hand, matching conditions at the boundary are involved in determining the size of the object \i.e its radius $R$ and also its total mass $M$. In this direction, some works available in the literature ~\citep{Farinelli2014,Capozziello2016} have argued that in the $f(R)$ domain the mass-radius profile undergoes modifications due to the presence of high order curvature terms such as $R^{2}$, $R^{3}$ etc. Besides, in ~\citep{Senovilla2013} was discussed the well-known Israel-Darmois ~\citep{Israel1966,Darmois1927} junction conditions in the framework of $f(R)$ gravity in considering both the isotropic and the anisotropic compact matter distributions.  They conclude that Israel-Darmois matching conditions are not satisfy at all in the $f(R)$ gravity arena. In considering $f(R,\mathcal{T})$ the situation becomes more complicated because we have geometrical and matter modifications. Notwithstanding, in this respect there is not geometrical contribution from the chosen model given by Eq. (\ref{frt}). This is so because, the geometrical part is just the Ricci's scalar and the matter contribution given by the trace of the energy momentum tensor $\mathcal{T}$, which is coupled via $\chi$ parameter, is the most simple one election. At this stage it is worth mentioning that the extra component $\mathcal{T}$ remains confined by the compact object. So, it vanishes beyond the stellar structure.\\
Based on the above explanation, the external space-time is described by the exterior Schwarzschild's solution given by
 \begin{eqnarray}\label{4.1}
&\qquad\hspace{-0.5cm} ds^{2} =-\frac{dr^{2}}{\left(1-\frac{2M}{r}\right)}+\left(1-\frac{2M}{r} \right) dt^{2} -r^{2} (d\theta ^{2}+\sin ^{2} \theta  d\phi ^{2} ).
\end{eqnarray}

To join (\ref{3.1})-(\ref{3.4}) with (\ref{4.1}) in a smoothly way, we need to ensure the fulfillment of the first and second fundamental forms. The first fundamental form requires the continuity of the metric potentials (intrinsic curvature) across the boundary $\Sigma=r=R$. Explicitly it reads
\begin{equation}
\left[ds^{2}\right]_{\Sigma}=0,
\end{equation}
\begin{equation}
e^{\lambda^{-}}|_{r=R}=e^{\lambda^{+}}|_{r=R} \quad \mbox{and} \quad   e^{\nu^{-}}|_{r=R}=e^{\nu^{+}}|_{r=R},
\end{equation}
and the second fundamental form is related with the continuity of the extrinsic curvature $K_{\mu\nu}$ induced by $\mathcal{M}^{-}$ and $\mathcal{M}^{+}$ on $\Sigma$. The continuity of the extrinsic curvature in ensured due to the absent of free material content which implies that the boundary $\Sigma$ is completely smooth and regular. Otherwise, the presence of thins shells or layers lead in this case a non continuous extrinsic curvature tensor across the surface ~\citep{Israel1966}. Nevertheless, as was pointed out earlier this is not the case and the boundary $\Sigma$ is completely regular. So, the continuity of $K_{rr}$ component across $\Sigma$ yields to
\begin{equation}\label{second}
{p}_{r}(R)=0.
\end{equation}
The above requirement (\ref{second}) determines the size of the object \i.e the radius $R$ which means that the material content is confined within the region $0\leq r\leq R$. Thus applying these conditions we arrive at the following equations by setting $m(r)=M$ as
\begin{eqnarray}\label{4.2}
 &&e^{\nu(R)}=C\,e^{2\,AR^2}=1-{\frac {2M}{R}},\\ \label{4.3}
 &&e^{\lambda(R)}= 1+D\,A{R}^{2}\, e^{2\,AR^2}= \left( 1-{\frac {2M}{R}} \right)^{-1}.
 \end{eqnarray}
Moreover, in order to ensure $p_{r}(R)=0$ at the boundary ~\citep{Lake1996,Lake2003,Lake2017} we have \begin{equation}
\label{4.4}
\nu^{\prime}(R)=\frac {2M}{R\,(R-2M)}.     
\end{equation}

 After solving the Eqs.~(\ref{4.2})-(\ref{4.4}) and using the relation $D=16\,C\,A\,B$, we obtain the values of different arbitrary constants and parameters as
 \begin{eqnarray}\label{eq39}
 & \qquad \hspace{-1.2cm} A ={\frac {M}{2\,{R}^{2} \left( R-2\,M\right) }},\\ \label{eq40}
 & \qquad C=\frac {\left(R-2M\right)}{R}\,e^{M/(2M-R)},\\ \label{eq41}
 &\qquad \hspace{-1.7cm} \qquad D= 4\,e^{M/(2M-R)},  \\  \label{eq42}
 & \qquad \hspace{-2.6cm} \qquad B={\frac {{R}^{3}}{2M}},
 \end{eqnarray}
 So, Eqs. (\ref{eq39})-(\ref{eq42}) are the necessary and sufficient equations to determine the complete set of constants parameters of the solution.

In order to discuss the complete structure of the stellar models in $f(R, \mathcal{T})$ gravity we need to determine the expressions of physical parameters like effective radial pressure $(p_r^{eff})$, tangential pressure $(p_t^{eff})$ and  effective energy density $({\rho}^{eff})$. For this purpose, we substitute the values of ${{\rm e}^{\nu}}$ from Eq.~(\ref{3.1}) and ${{\rm e}^{\lambda}}$ from Eq.~(\ref{3.4}) into Eqs.~(\ref{2.3})-(\ref{2.2}) and using EOS (\ref{2.5}) with Eqs.(\ref{2.4a}-\ref{2.2a}) we obtain the expression of the effective density~$\left({\rho}^{{\it eff}}\right)$, the effective radial pressure~$\left(p_r^{{\it eff}}\right)$ and the effective tangential pressure~$\left(p_t^{{\it eff}}\right)$ as
\begin{eqnarray} \label{3.6}
&\qquad\hspace{-3.0cm}\resizebox{0.9\hsize}{!}{$ p_r^{{\it eff}}=\frac{1}{24\,\pi\,\,(4\,\pi+\chi)\left[-2\, f(r)\,M\,r^2 + (2\,M-R)\,R^2\right]^2}\,\Big[-12\,B\,\chi^2\,\left[-2\, f(r)\,M\,r^2 + (2\,M-R)\,R^2\right]^2-6\,\pi\,\Big(64\,B\, f(r)\,M^2\,\pi\,r^4$} \nonumber\\
&\qquad\hspace{0.0cm}+ \resizebox{0.9\hsize}{!}{$(2\,M-R)\,R^2\,(M + 32\,B\,M\,\pi\,R^2 -16\,B\,\pi\,R^3)\Big) +12\,\pi \,f(r)\,M\,\left[(1 - 32\,B\,\pi\, r^2)\,R^3 + 2\,M\,(r^2 - R^2 + 32\,B\,\pi\,r^2 R^2)\right]$}\nonumber\\
&\qquad\hspace{-0.5cm}\resizebox{0.9\hsize}{!}{$+\chi\,\left(-72\,B\,\pi\, R^6+288\,B\,M\,\pi\,R^3\,[-f(r)\,r^2 + R^2]\right) +\chi\,M^2\,\Big(-288\,B\,f^2(r)\,\pi\, r^4 -288\,B\,\pi\,R^4 +r^2\,[1 + 4\,f^2(r)$} \nonumber\\
&\qquad\hspace{11.0cm} +f(r)\, (-4 + 576\,B\,\pi\,R^2)]\Big) \Big],~~~
\end{eqnarray}

\begin{eqnarray} \label{3.7}
 &\qquad\hspace{-3.cm} \resizebox{0.90\hsize}{!}{$p_t^{{\it eff}}=-\frac{1}{24\,\pi\,\,(4\,\pi+\chi)\left[-2\, f(r)\,M\,r^2 + (2\,M-R)\,R^2\right]^2}\,\Big[12\,B\,\chi^2\,\left[-2\, f(r)\,M\,r^2 + (2\,M-R)\,R^2\right]^2+4\,\chi\,\Big[18\,B\,\pi\,R^6-72\,B\,M\,R^3 $}\nonumber\\
  &\qquad\hspace{-1.5cm}\resizebox{0.90\hsize}{!}{$ (-r^2\,f(r)+R^2)+M^2\,\Big(72\,B\,f(r)\,\pi\,r^4+72\,B\,\pi\,R^4-r^2\left[1+4\,f^2(r)+4\,f(r)\,(-1+36\,B\,\pi\,R^2)\right]\Big)\Big]$}\nonumber\\
 &\qquad\hspace{-1.cm}
 \resizebox{0.90\hsize}{!}{$+6\,\pi\,\Big[16\,B\,\pi\,R^6 - M\,R^3\,\left(1 + f(r)\, (2 - 64\,B\,\pi\,r^2) +64\,B\,\pi\,R^2)\, +\right)\Big]+4\,M^2\,\big[32\,B\,f^2(r)\,\pi\,r^4+R^2(1+2\,f(r)$}\nonumber\\
 &\qquad\hspace{6.0cm}+32\,B\,\pi\,R^2)-r^2\,\big] -6\,\pi\,r^2\,\left(1 + 4\,f^2(r) +  f(r) (-2 + 64\, B\,\pi\, R^2)\right)\Big],~~~~
\end{eqnarray}

\begin{eqnarray}\label{2.5a}
&\qquad\hspace{-2.0cm} {\rho}^{{\it eff}}= \frac{1}{24\,\pi\,\,(4\,\pi+\chi)\left[-2\, f(r)\,M\,r^2 + (2\,M-R)\,R^2\right]^2}\,\Big[-12\,B\,\chi^2\,\left[-2\, f(r)\,M\,r^2 + (2\,M-R)\,R^2\right]^2+6\,\pi\,\Big(64\,B\, f(r)\,M^2\,\pi\,r^4 \nonumber\\
&\qquad\hspace{-0.50cm}+ (2\,M-R)\,R^2\,(-3\,M + 32\,B\,M\,\pi\,R^2 -16\,B\,\pi\,R^3)\Big) +12\,\pi \,f(r)\,M\,\big[-(3 + 32\,B\,\pi\, r^2)\,R^3 + M\,(6\,R^2 +r^2 \nonumber\\
&\qquad\hspace{0.0cm}+ 64\,B\,\pi\,r^2 R^2)\big]+\chi\,\left(72\,B\,\pi\, R^6-6\,M\,R^3\,[-1-2\,f(r)(1+24\,B\,\pi\,r^2)+48 B\,\pi\, R^2]\right) +\chi\,M^2\,\Big(288\,B\,f^2(r)\,\pi\, r^4 \nonumber\\
&\qquad\hspace{2.5cm}-12\,R^2\,[1 + 2\,f^2(r) -24\,B\,\pi\,R^2]-r^2[1+4\,f^2(r)+4\,f(r)\,(144\,B\,\pi\,R^2-7)]\Big) \Big].
\end{eqnarray}

where,~~~ $f(r)=\exp\left[\frac{M (R^2-r^2)}{(2 M - R) R^2} \right]$

%%%%%%%%%%%%%%%%%%%%%%%%%%%%%%%%%%
\begin{figure}[h!]
\begin{center}
\includegraphics[width=5.5cm]{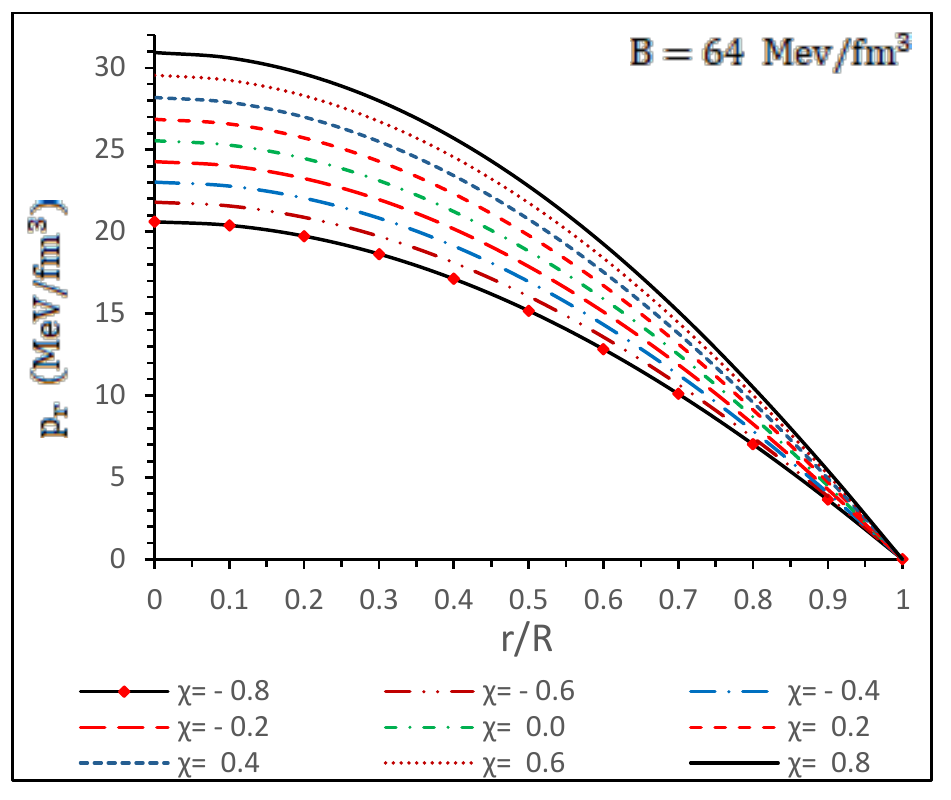}
\includegraphics[width=5.5cm]{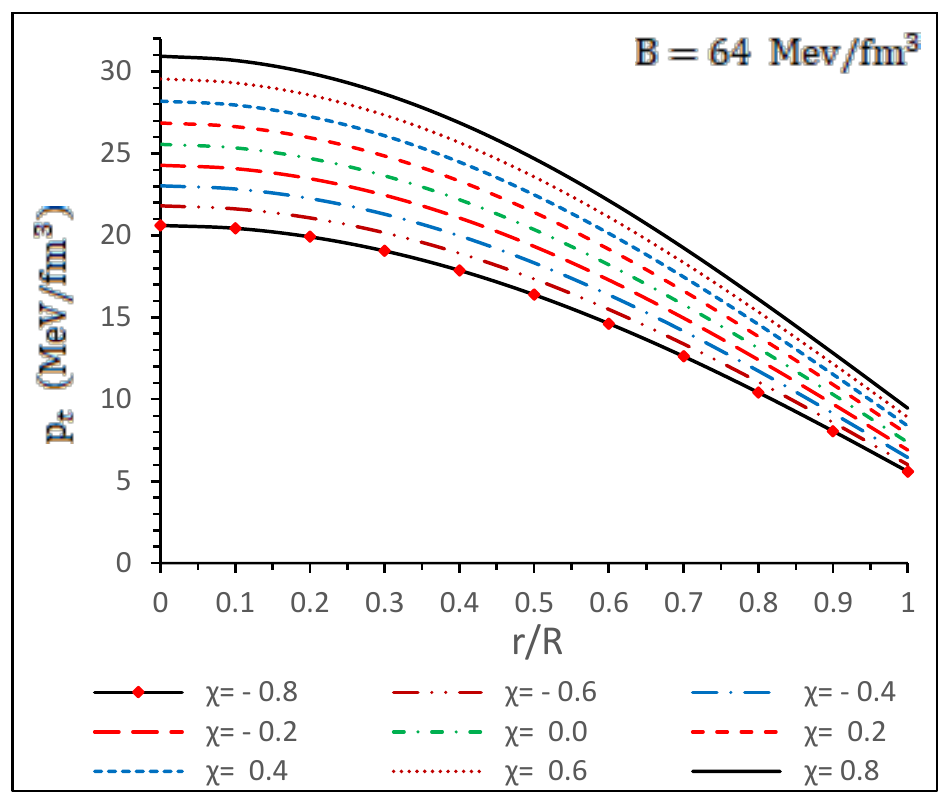}
\includegraphics[width=5.5cm]{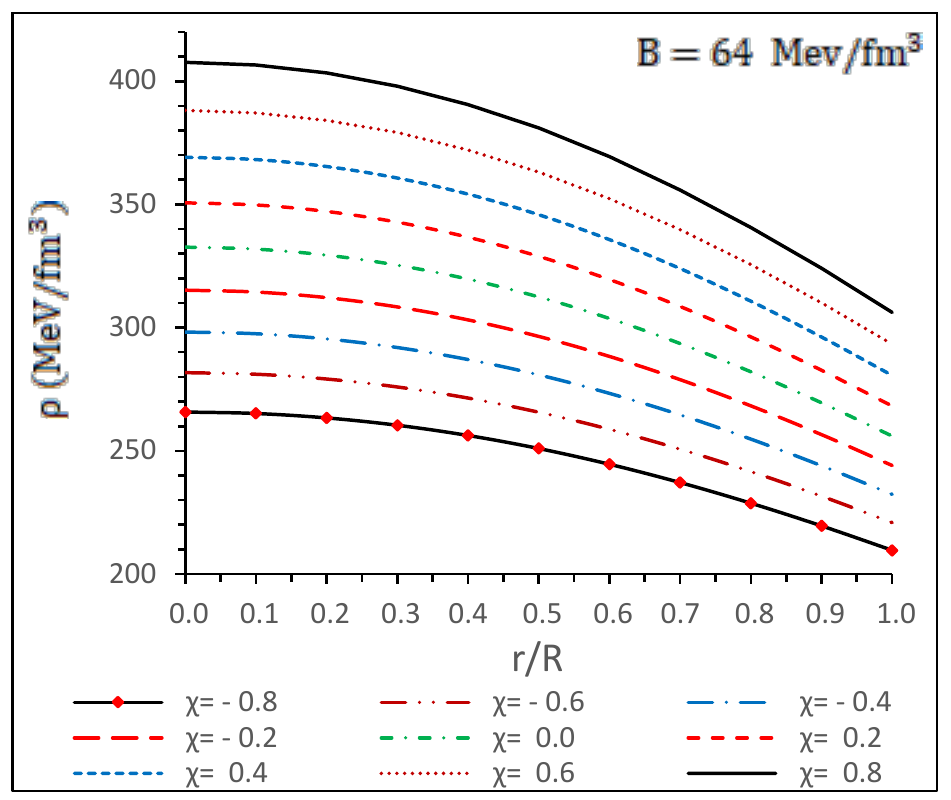}
\includegraphics[width=5.5cm]{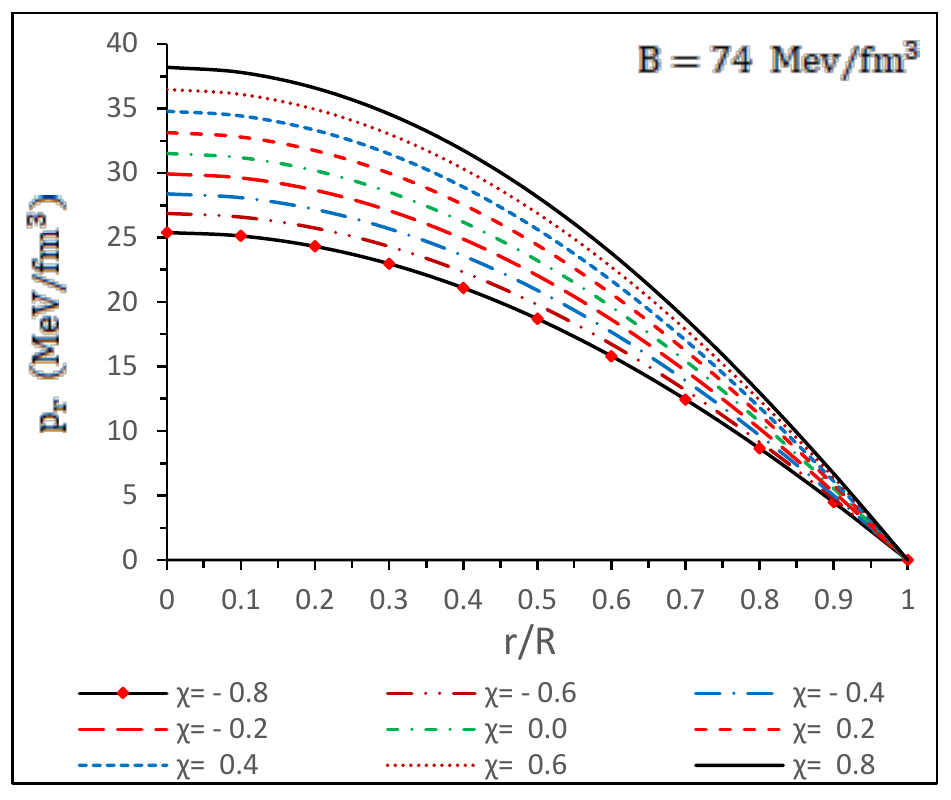}
\includegraphics[width=5.5cm]{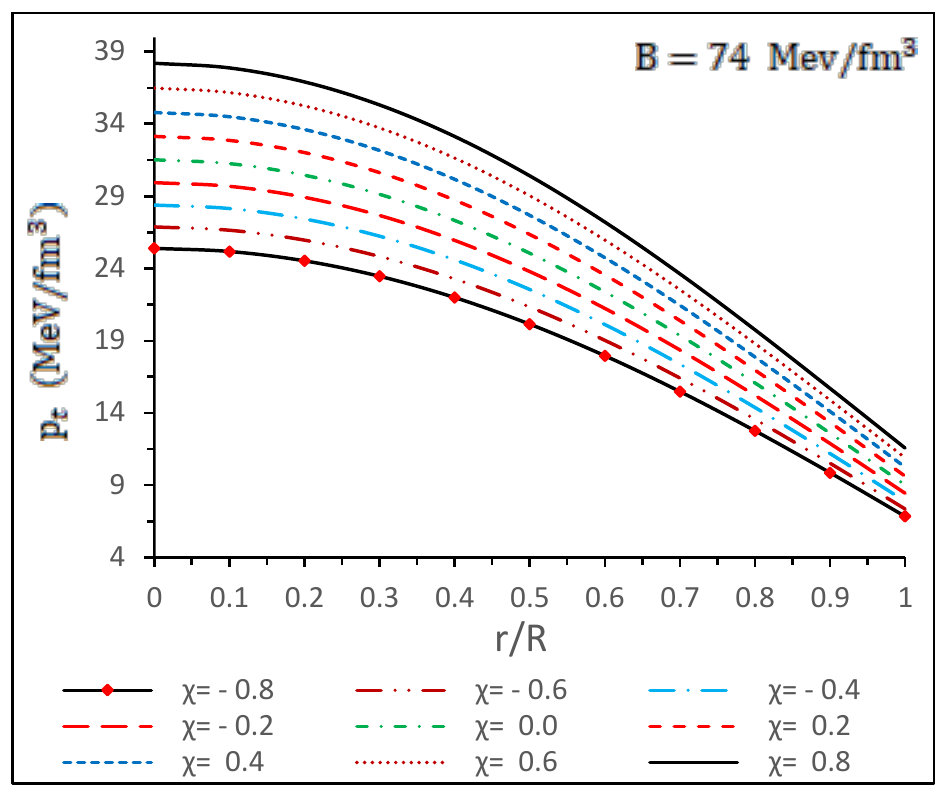}
\includegraphics[width=5.5cm]{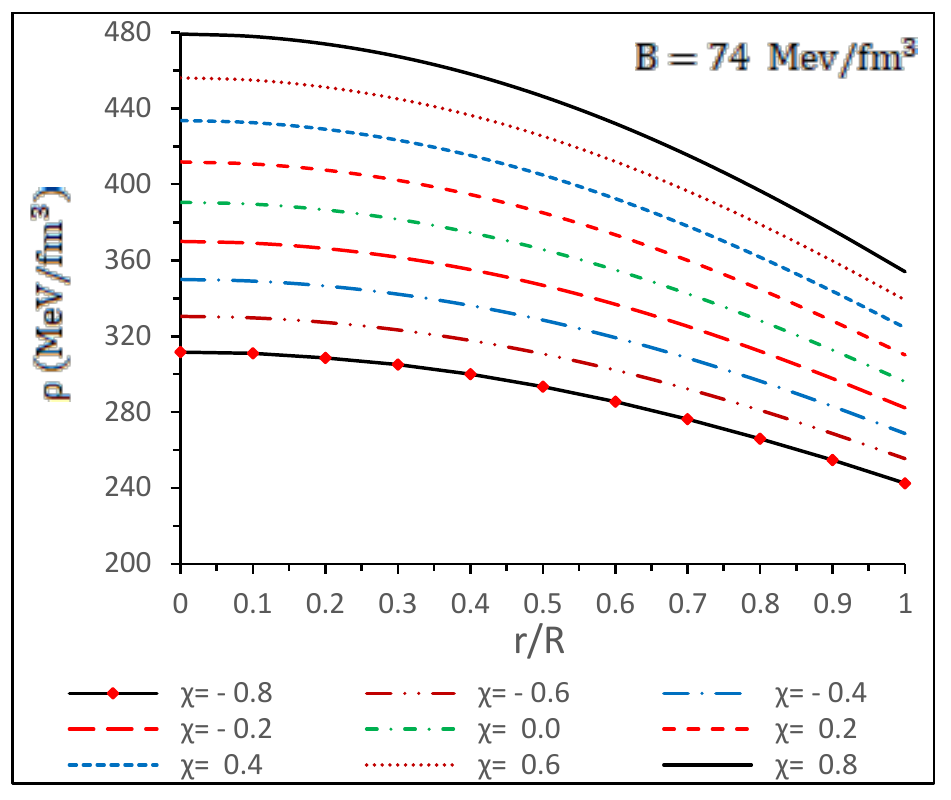}
\caption{Behavior of effective radial pressure ($p_r^{eff}$), effective tangential pressure ($p_t^{eff}$) and effective energy density ($\rho^{eff}$) vs. radial coordinates $r/R$ of SMC X-4 for different values of $\chi$ with bag constant $B=64 Mev/fm^3$ and $B=74 Mev/fm^3$.}\label{Fig2}
\end{center}
\end{figure}
%%%%%%%%%%%%%%%%%%%%%%%%%%%%%%%%%%%%%%%

\begin{figure}[h!]
\begin{center}
\includegraphics[width=7cm]{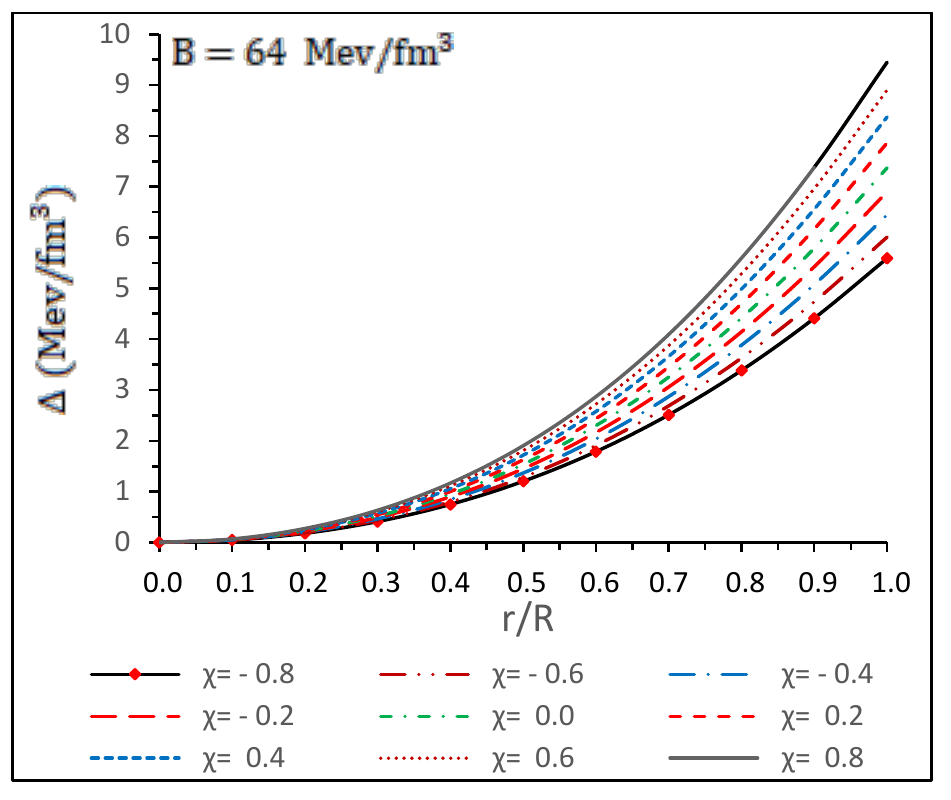}
\includegraphics[width=7cm]{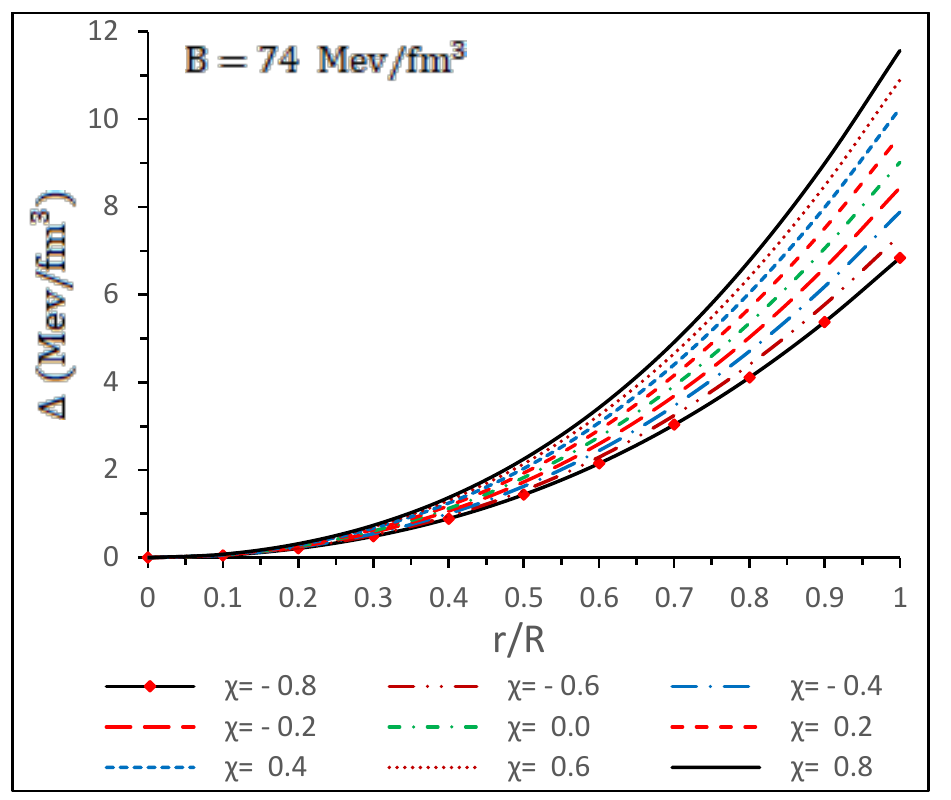}
\caption{Behavior of effective anisotropy $(\Delta_{eff})$ vs. radial coordinates $r/R$ of SMC X-4 for different values of $\chi$ with bag constant $B=64 Mev/fm^3$ and $B=74 Mev/fm^3$.}\label{Fig3}
\end{center}
\end{figure}

The profile of physical quantities such as, effective radial pressure $(p_r^{{\it eff}})$, effective tangential pressure $(p_t^{{\it eff}})$ and effective energy density $({\rho}^{{\it eff}})$ as a function to the radial coordinate $r$ are presented in Figs.~\ref{Fig2} and~\ref{Fig3}. From these figures, we discover that the three physical quantities are maximally at the origin and decrease monotonously to achieve their minimum values at the surface, which proves the physical availability of the anticipated stellar model. These figures likewise highlight that the energy density and the tangential and radial pressures are positive and regular at the origin, which proves that our framework is free from physical and mathematical singularities.

Due to the existence of tangential and radial pressures in the structure, the anisotropy of the astrophysical structure, in $f(R,\mathcal{T})$ gravity is acquired by utilizing Eqs.~(\ref{3.6}) and (\ref{3.7}) as follows

 \begin{eqnarray}\label{3.8}
&\qquad\hspace{-5.15cm} \Delta_{eff} = \frac{\left[1-2\,f(r)\right]^2\,M^2\,r^2}{8\,\pi\,\Big(2\,f(r)\,M\,r^2+R^2\,(R-2\,M)\Big)^2}. \label{Deff2}
\end{eqnarray}
%%%%%%%%%%%%%%%%%%%%%%%%%%%%%%%%%%
%%%%%%%%%%%%%%%%%%%%%%%%%%%%%%%%%%%%%%%

%%%%%%%%%%%%%%%%%%%%%%%%%%%%%%%%%%
\begin{figure}[h!]
\begin{center}
\includegraphics[width=8.2cm]{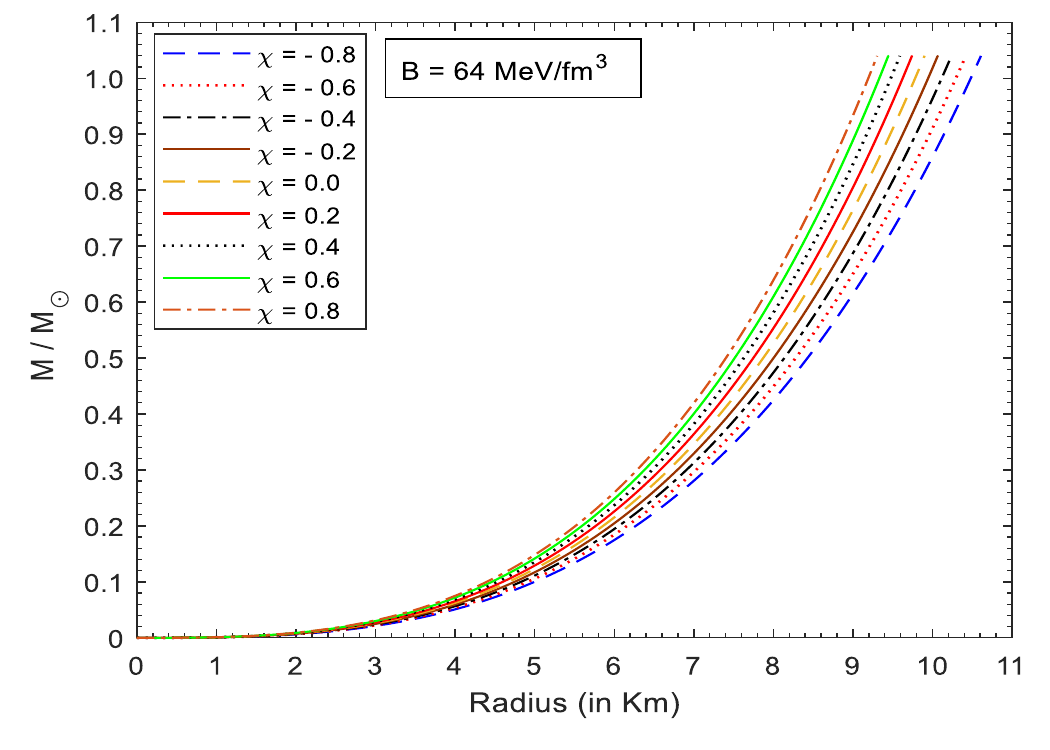} \includegraphics[width=8.2cm]{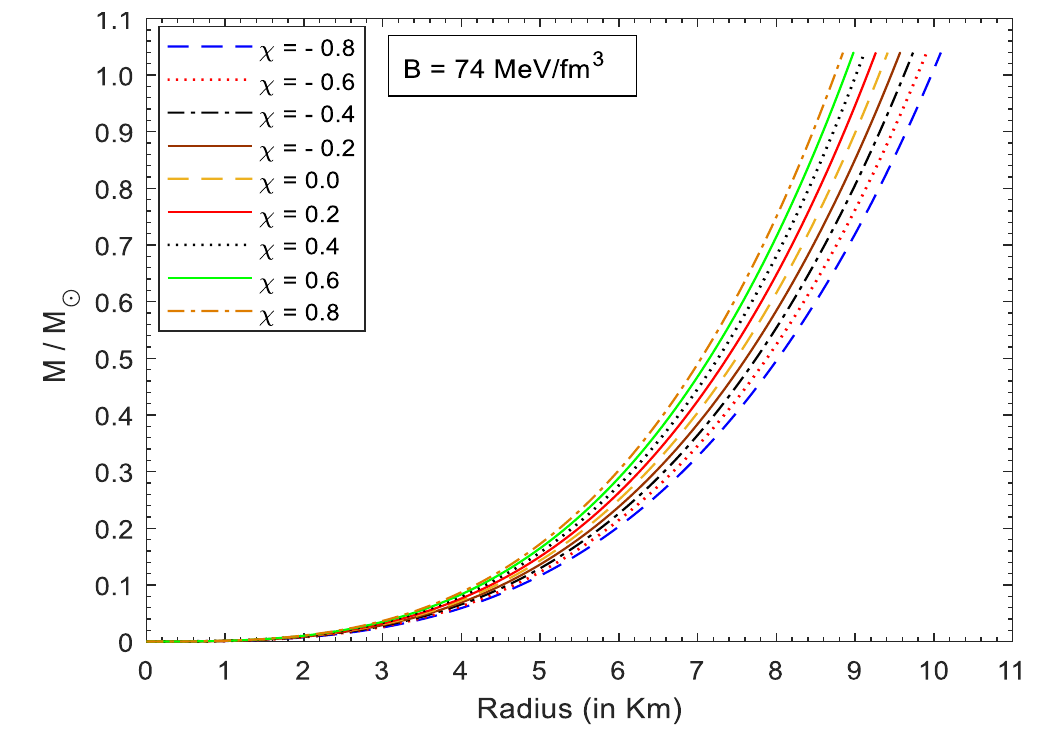}
\caption{Behavior of total mass ($M/M_{\odot}$) verses  radius (R) of compact star $SMC X-1$ for different $\chi$ with bag constant $B=64 MeV/fm^3$ and $B=74 MeV/fm^3$.}\label{Fig.6}
\end{center}
\end{figure}
%%%%%%%%%%%%%%%%%%%%%%%%%%%%%%%%%%%%%%%

We use Eq.~(\ref{1.4}), in the context of the modified $f\left(R,\mathcal{T} \right)$ gravity theory, in order to obtain the modified form of the stress-energy tensor expressed by Eq.~(\ref{1.6}), in a more explicit form as follows:

\begin{eqnarray}\label{4.9a}
 -\frac{\nu^{{\prime}}}{2}\, \left( \rho+p_r \right)-\frac{dp_r}{dr}+\frac{2}{r}\left({p_t}-{p_r}\right) +{\frac {\chi}{3(8\pi +2\chi)}}\left(3\,\frac{d\rho}{dr}-\frac{dp_r}{dr}-2\,\frac{dp_t}{dr} \right)=0.
\end{eqnarray}

Therefore, the equilibrium of hydrostatic for the anisotropic spherically symmetric compact astrophysical structure can be assured with the $f\left(R,\mathcal{T} \right)$ model by the use of Eqs.~(\ref{2.3}), (\ref{2.8}), and (\ref{4.9a}) as follows:

\begin{eqnarray}\label{4.10}
 \frac{dp_r}{dr}=\frac{12\,(p_t -p_r)\,\left(\,1- \frac{2m}{r}\,\right) + (\rho + p_r)\, \left[-6\,\frac{m}{r} + r^2\,(3\,\rho\,\chi - 7\,\chi\,p_r - 2\,\chi\,p_t - 24\,\pi\, p_r\,)\,\right]}{6\,(r -2\,m)\,\left[1- \frac{\chi}{6\,(\chi +4 \pi)}\, \left(3\,\frac{d\rho}{dp_r} - 1 - 2\,\frac{dp_t}{dp_r}\right)\right]}.
\end{eqnarray}

 By using the Eqs.(\ref{second}), (\ref{3.6}) and (\ref{eq39}-\ref{eq42}) we obtain total mass $M$ of the strange star $R$ as,
\begin{equation}
 M=\frac{R}{2\left(\chi - 12 \pi\right)}\left[-18\pi + \sqrt{324\pi^2+4R^2(\chi - 12\pi)(12B\chi^2+ 72B\chi\pi+ 96B\pi^2)}\right].\label{Mass1}
\end{equation}
It is worth mentioning that the total mass $m(R)=M$ expressed by the above equation (\ref{Mass1}) comes from the continuity of the remaining components of the extrinsic curvature across the surface $\Sigma$ \i.e, $K_{\theta\theta}$ and $K_{\phi\phi}$ ~\citep{Lake2017}.
In order to calculate the exact values of the radius of the compact anisotropic stellar structures at the different values of $\chi$, we use the Eqs. (\ref{2.5}) and (\ref{Mass1}) to solve the hydrostatic equation (\ref{4.10}). To this end, we take into account the observed values for the mass of strange star candidates and take some specific values for $B$ and $\chi$. We find standard hydrostatic equation for compact anisotropic stellar structures in Einstein gravity by inserting the specific value of $\chi=0$ in Eq. (\ref{4.10}).

\begin{table*}
  \centering
    \caption{Physical properties of SMC~X-1 due to different values of $\chi$ for $B=64 MeV/fm^3$.}\label{Table 3}
       \scalebox{0.95}{
\begin{tabular}{ ccccccccccccccccccccccccccc}
\hline
Values  & Predicted     &  ${\rho}^{eff}_c$ &  ${\rho}^{eff}_s$ &  $p_c$ & $\frac{2M}{R}$ & $Z_s$ & $A$ & $C$ & $D$\\
of $\chi$  & radius $(km)$ &   $(gm/{cm}^3)$   & $(gm/{cm}^3)$     & $(dyne/{cm}^2)$ & & & $(km^{-2})$ &  &\\ \hline

-0.8 & $10.6096^{+0.274}_{-0.292}$ &  $4.74034 \times 10^{14}$ &	$3.73971 \times 10^{14}$ & $3.30874 \times 10^{34}$ &0.28917 & 0.18609&  $9.0352\times 10^{-4}$&0.57999&3.26378\\

-0.6 & $10.42^{+0.269}_{-0.287}$ & $5.02548 \times 10^{14}$ & $3.93991 \times 10^{14}$ & $3.50073 \times 10^{34}$ & 0.29443 & 0.19051& $9.6085\times 10^{-4}$&0.57269&3.24671\\

-0.4 & $10.2392^{+0.274}_{-0.282}$ &  $5.31941 \times 10^{14}$ &	$4.14442 \times 10^{14}$ & $3.69724 \times 10^{34}$ &0.29964 & 0.19492 & 0.0010202&0.56548& 3.22965\\

-0.2  & $10.0664^{+0.259}_{-0.277}$ & $ 5.62216 \times 10^{14}$ & $4.35324 \times 10^{14}$ & $3.89809 \times 10^{34}$ &0.30478 & 0.19933& 0.0010815& 0.55837 & 3.21263\\

0 & $9.901^{+0.255}_{-0.272}$ &  $5.93380 \times 10^{14}$ & $4.56634 \times 10^{14}$ & $4.10322 \times 10^{34}$ & 0.309869 & 0.20374 &0.001145&0.55135&3.19566\\

0.2 & $9.7426^{+0.250}_{-0.267}$ & $6.25446 \times 10^{14}$ & $4.78371 \times 10^{14}$ & $4.31272 \times 10^{34}$ & 0.31491& 0.20816&0.0012106 &0.544424&3.17868\\
0.4 & $9.5907^{+0.246}_{-0.263}$ &  $6.58422 \times 10^{14}$ & $5.00531 \times 10^{14}$ & $4.52656 \times 10^{34}$ & 0.31989& 0.21258 &0.001278 &0.53757 &3.16172\\
0.6 & $9.445^{+0.242}_{-0.259}$ & $6.92322 \times 10^{14}$ & $5.23113 \times 10^{14}$ & $4.74490 \times 10^{34}$ & 0.32484& 0.21702&0.0013484&0.53079 & 3.14471\\
0.8 & $9.3047^{+0.238}_{-0.254}$ &  $7.27134 \times 10^{14}$ & $5.46115 \times 10^{14}$ & $4.96711 \times 10^{34}$ & 0.32973& 0.22145&0.0014205&0.524114&3.12778\\
\hline
\end{tabular}}
  \end{table*}

\begin{table*}[h!]
  \centering
    \caption{Physical properties of SMC~X-1 due to different values of $\chi$ for $B=74 MeV/fm^3$.}\label{Table 4}
       \scalebox{0.95}{
\begin{tabular}{ ccccccccccccccccccccccccccc}
\hline
Values  & Predicted     &  ${\rho}^{eff}_c$ &  ${\rho}^{eff}_s$ &  $p_c$ & $\frac{2M}{R}$ & $Z_s$ & $A$ & $C$ & $D$\\
of $\chi$  & radius $(Km)$ &   $(gm/{cm}^3)$   & $(gm/{cm}^3)$     & $(dyne/{cm}^2)$ & & & $(km^{-2})$ &  &\\ \hline

-0.8 & $10.0894^{+0.259}_{-0.277}$ &  $5.55756 \times 10^{14}$ &	$4.32461 \times 10^{14}$ & $4.07674 \times 10^{34}$ &0.30408 & 0.19873&  0.0010731&0.55933&3.21496\\

-0.6 & $9.909^{+0.254}_{-0.272}$ & $5.89405 \times 10^{14}$ & $4.55597 \times 10^{14}$ & $4.31481 \times 10^{34}$ & 0.30964 & 0.20354 &0.001142&0.551673&3.19643\\

-0.4 & $9.7366^{+0.250}_{-0.267}$ &  $6.24092 \times 10^{14} $ &	$4.79231 \times 10^{14}$ & $4.55808 \times 10^{34}$ & 0.315107 & 0.208339 & 0.0012133&0.544147& 3.178002\\

-0.2  & $9.572^{+0.245}_{-0.262}$ & $ 6.59844 \times 10^{14}$ & $5.03361 \times 10^{14}$ & $4.80705 \times 10^{34}$ &0.32052 & 0.213146& 0.001287& 0.536711 & 3.15956 \\

0 & $9.4145^{+0.245}_{-0.262}$ &  $6.96665 \times 10^{14}$ & $5.279835 \times 10^{14}$ & $5.06149 \times 10^{34}$ & 0.325882 & 0.217956 &0.0013635&0.52937&3.141137\\

0.2 & $9.2637^{+0.237}_{-0.253}$ & $7.345713 \times 10^{14}$ & $5.530962 \times 10^{14}$ & $5.32152\times 10^{34}$ & 0.33119 & 0.22278 & 0.0014426 &0.522122&3.122698\\

0.4 & $9.119^{+0.232}_{-0.249}$ & $ 7.7356528 \times 10^{14}$ & $5.78696 \times 10^{14}$ & $5.58688 \times 10^{34}$ & 0.33644 & 0.22761&0.001524&0.514968 & 3.10428\\

0.6 & $8.9802^{+0.229}_{-0.245}$ &  $8.136703 \times 10^{14}$ & $6.04782 \times 10^{14}$ & $5.85791 \times 10^{34}$ & 0.34165 & 0.232459 &0.001609 &0.5078851 &3.085818\\

0.8 & $8.8467^{+0.225}_{-0.241}$ &  $8.54882 \times 10^{14}$ & $6.31349 \times 10^{14}$ & $6.134139 \times 10^{34}$ & 0.34681 & 0.23731 &0.001696&0.500896&3.067375\\

\hline
\end{tabular}  }
  \end{table*}

\begin{table*} [h!]
  \centering
    \caption{Physical properties of $SMC~X-1$ due to different values of $B$ for $\chi=0.2$.}\label{Table 5}
      \scalebox{1.0}{
\begin{tabular}{ ccccccccccccccccccccccccccc}
\hline
Values  & Predicted  &  ${\rho}^{eff}_c$ &  ${\rho}^{eff}_0$ &  $p_c$ & $\frac{2M}{R}$ & $Z_s$  \\
of $B$ & radius $(Km)$&  $(gm/{cm}^3)$ & $(gm/{cm}^3)$ & $dyne/{cm}^2$ & & \\
\hline
56  & $10.205^{+0.262}_{-0.282}$ & $5.3995509 \times 10^{14}$ &  $ 4.1858744 \times 10^{14}$ & $3.558838 \times 10^{34}$ &  0.30074 & 0.19586\\

86 & $8.7917^{+0.223}_{-0.239}$ & $8.686234 \times 10^{14}$ &  $6.42762 \times 10^{14}$ & $6.62323 \times 10^{34}$ & 0.34901 & 0.239401\\

96 & $8.461^{+0.213}_{-0.229}$ & $9.82751 \times 10^{15}$ &  $7.17479 \times 10^{14}$ & $7.77902 \times 10^{35}$ & 0.36267 & 0.25262\\

\hline
\end{tabular}  }
  \end{table*}

 \section{Physical properties of the anisotropic astrophysical structure in $f(R,\mathcal{T})$ gravity theory}\label{sec5}
We are presently in a situation to test physical highlights of the anisotropic stellar structure in framework of $f\left(R,\mathcal{T}\right)$ gravity theory, in order to study the energy conditions,compactness and gravitational surface redshift, the status of the sound speed within the stellar system, the Modified TOV equation, the adiabic index, etc., in the following subsections.

\subsection{Energy conditions of the stellar model in the $f\left(R,\mathcal{T}\right)$ theory gravity}\label{subsec5.1}
For the mathematical functions of stress-energy tensors to represent physically reasonable matter fields, certain special constraints, widely known as energy conditions, must be respected. Chakraborty~\citep{SC2013} has been presented a formalism to check the validity of the energy conditions in the framework of modified gravity. For our choice of $f\left(R,\mathcal{T}\right)$ gravity theory written in the specific form $f\left(R,\mathcal{T}\right)=R+2\chi\mathcal{T}$, having an effective density, an effective radial pressure, and an effective tangential pressure, the Weak Energy Condition (WEC), the Null Energy Condition (NEC), the Strong Energy Condition (SEC), the Dominant Energy Condition (DEC) and Trace energy condition (TEC) are formulated as follows:
 {\small{\begin{eqnarray}\label{5.1.1}
&\qquad\hspace{-5.3cm} ~NEC:~~{{\rho}^{eff}}\geq 0,\\ \label{5.1.2}
&\qquad\hspace{-1.8cm} ~WEC:~~{{\rho}^{eff}}+{{p^{eff}_r}}\geq 0,~{{\rho}^{eff}}+{{p^{eff}_t}}\geq 0, \\ \label{5.1.3}
&\qquad\hspace{-3.3cm} ~SEC:~{\rho}^{eff}+{p^{eff}_r}+2{p^{eff}_t} \geq 0, \\ \label{5.1.4}
&\qquad\hspace{-2.2cm}  ~DEC:~{\rho}^{eff}-{p^{eff}_r} \geq 0,~{\rho}^{eff}-{p^{eff}_t} \geq 0 \\  \label{5.1.4}
&\qquad\hspace{-3.5cm}  ~TEC:~{\rho}^{eff} - {p^{eff}_r} -2\,{p^{eff}_t} \geq 0
 \end{eqnarray}}}

 %%%%%%%%%%%%%%%%%%%%%%%%%%%%%%%%%%
\begin{figure}[h!]
\begin{center}
\includegraphics[width=4cm]{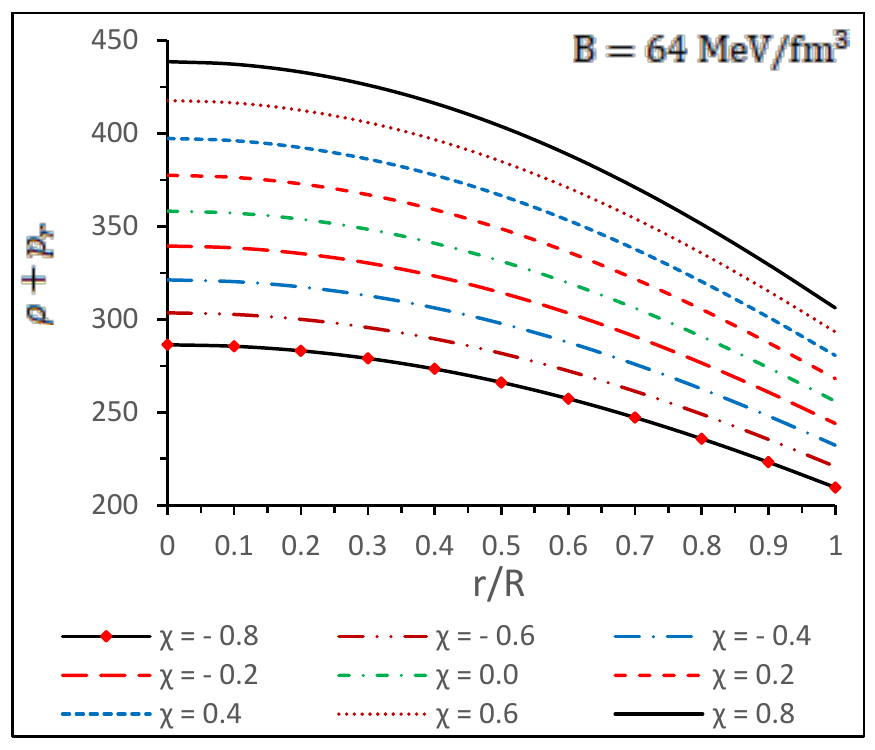}
\includegraphics[width=4cm]{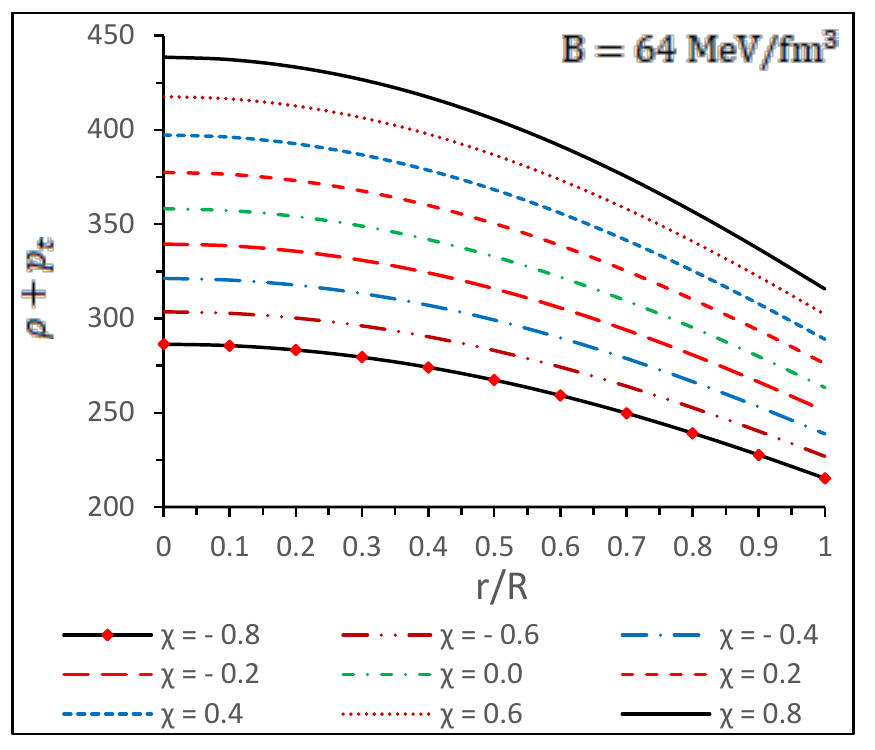}
\includegraphics[width=4cm]{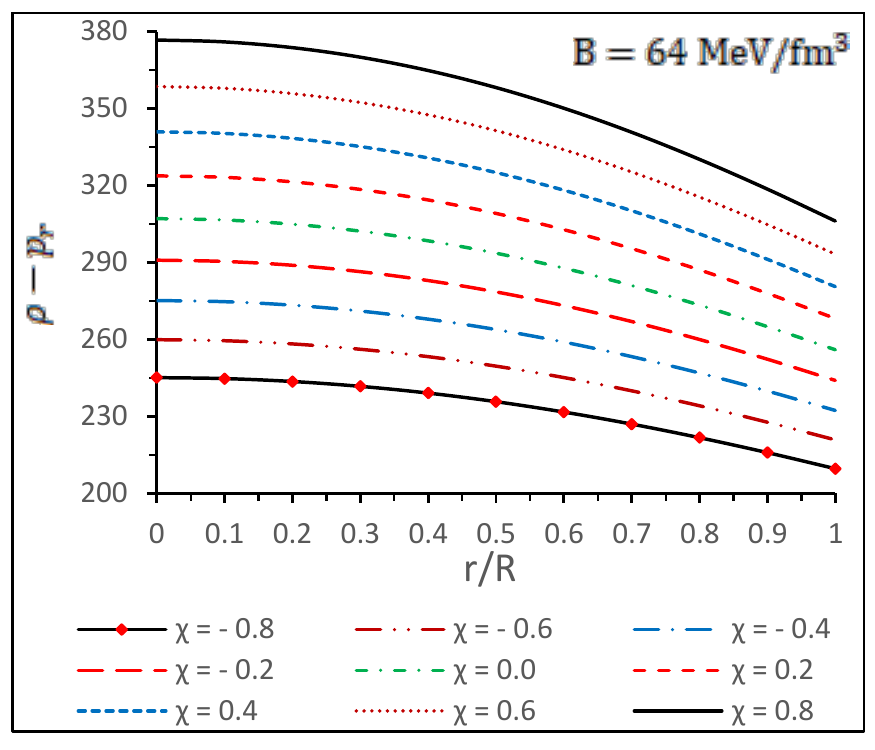}
\includegraphics[width=4cm]{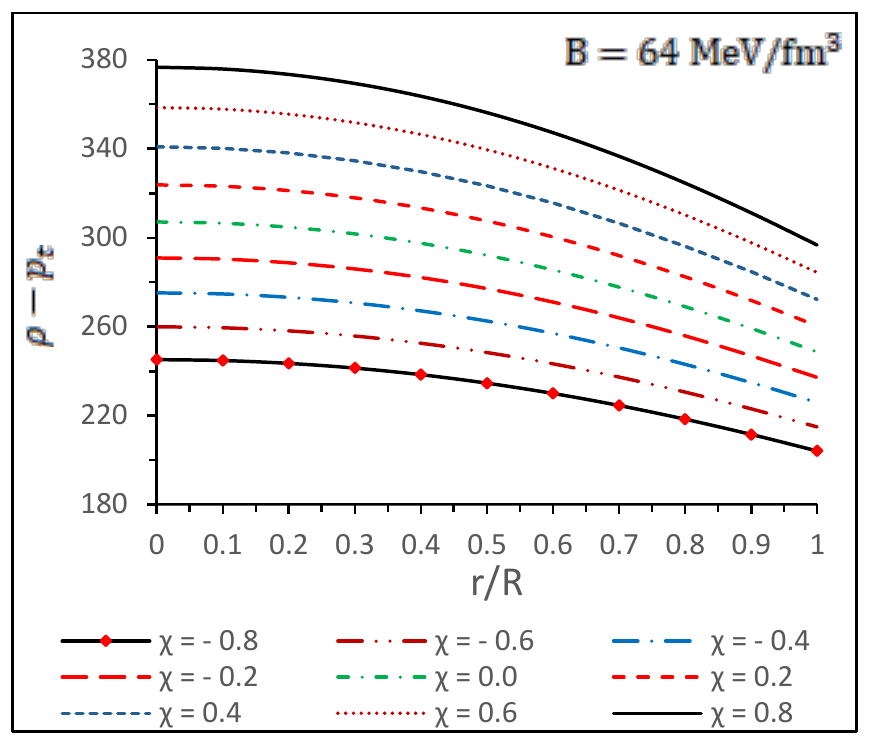}
\includegraphics[width=4cm]{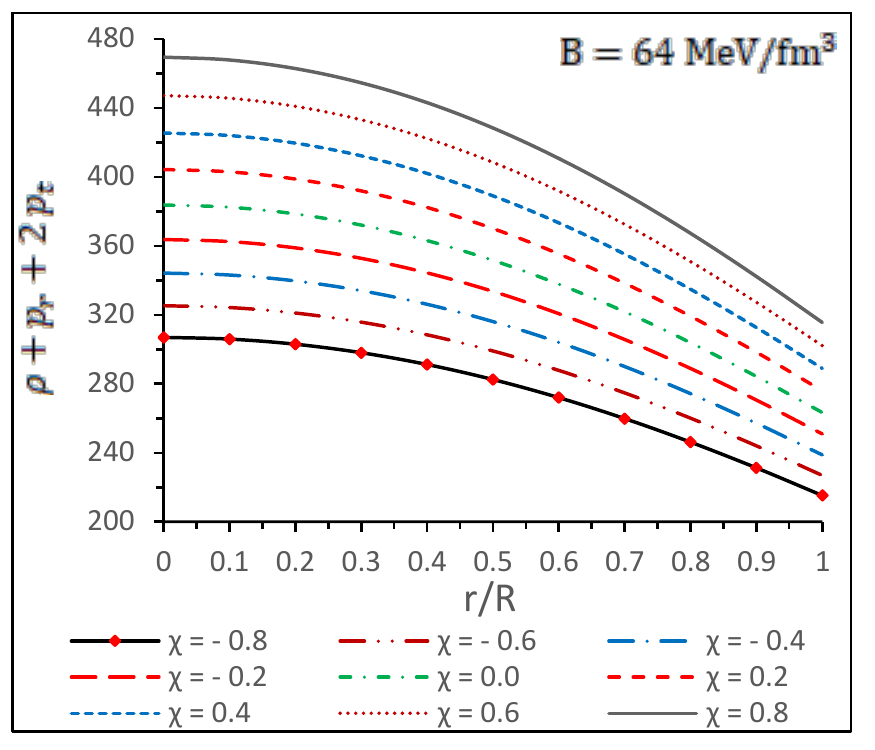}
\includegraphics[width=4cm]{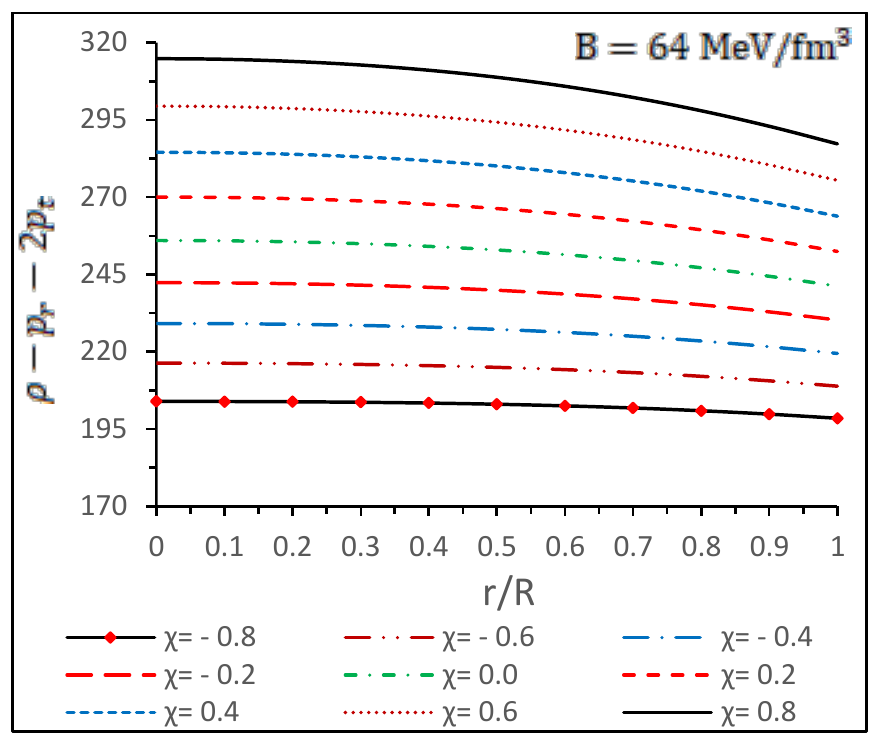}
\includegraphics[width=4cm]{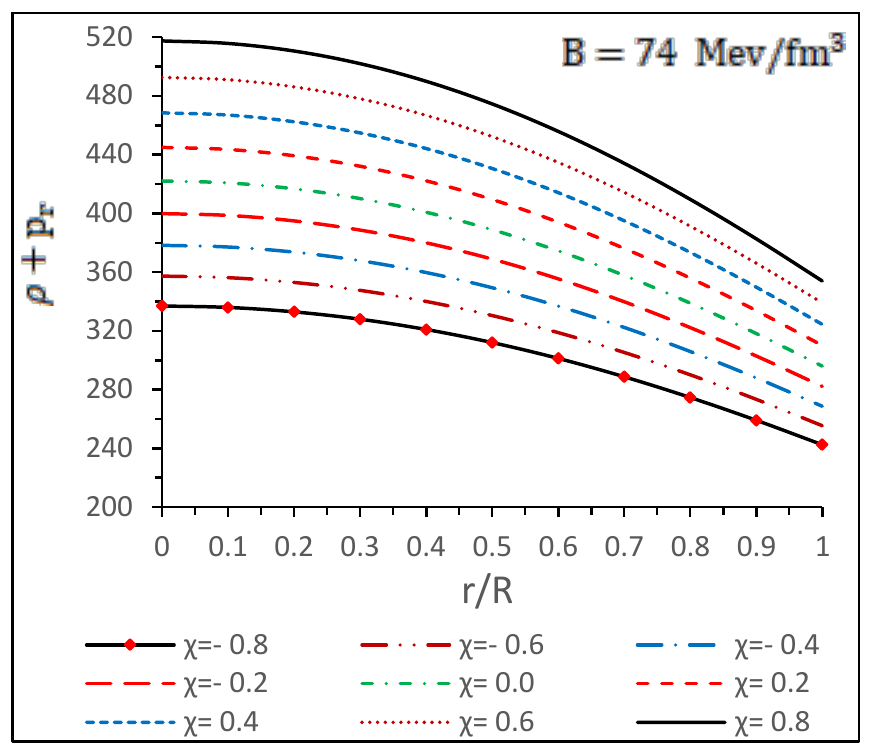}
\includegraphics[width=4cm]{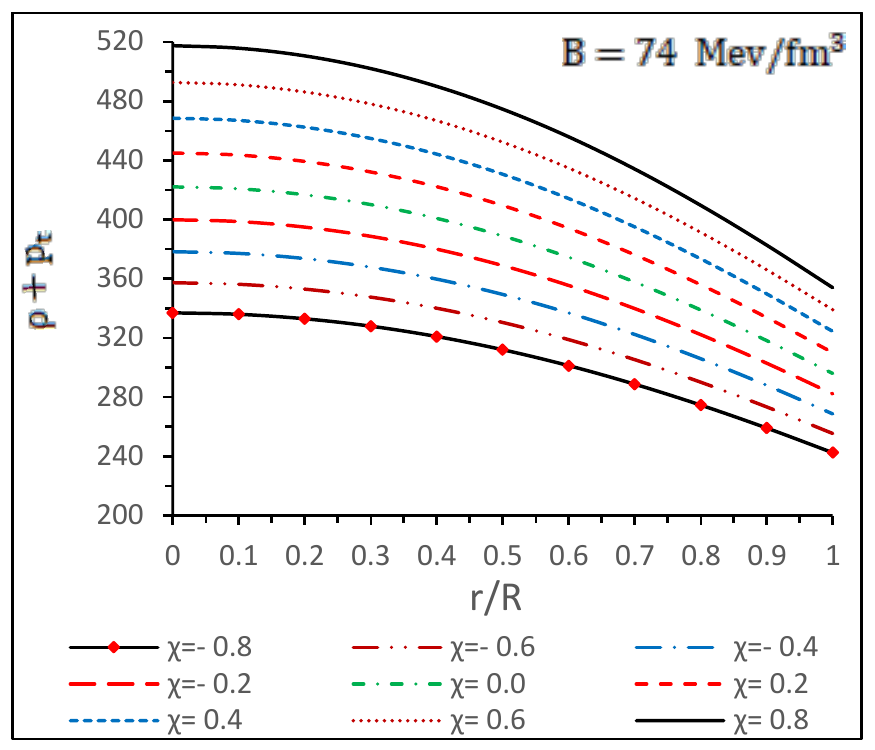}
\includegraphics[width=4cm]{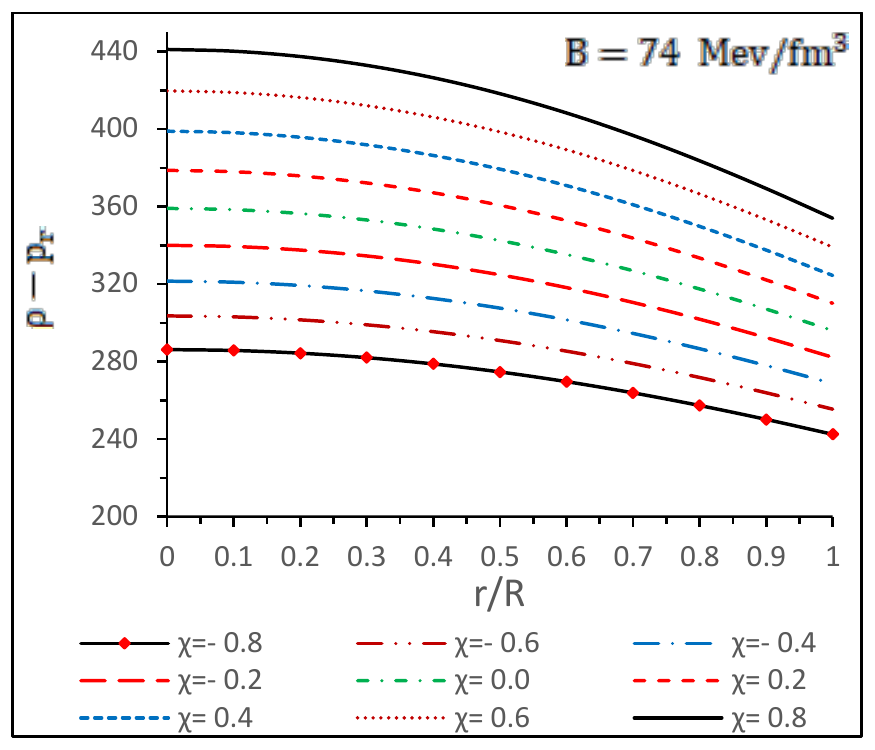}
\includegraphics[width=4cm]{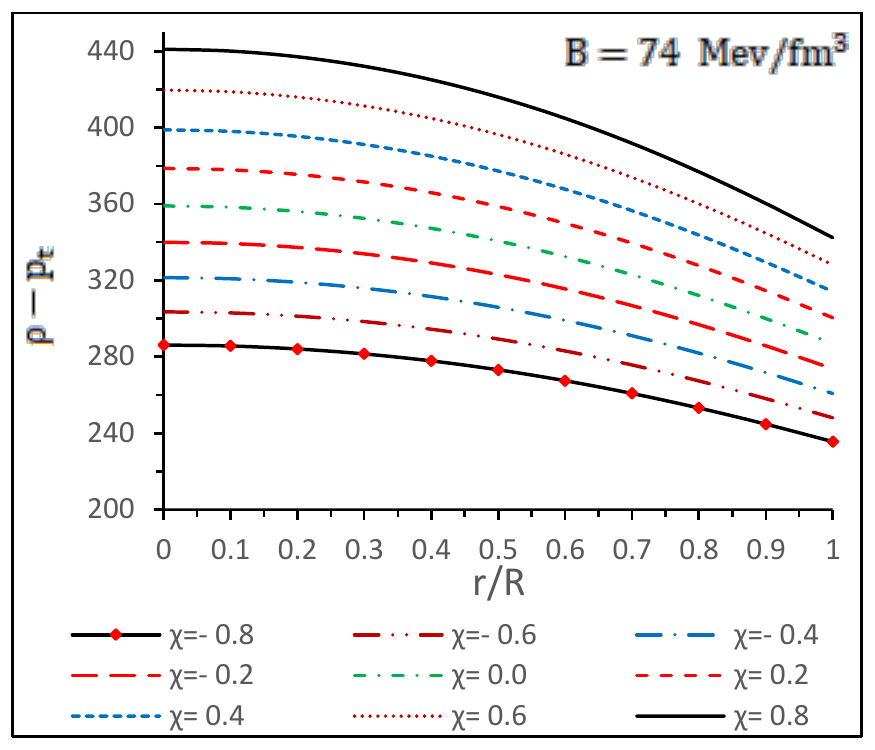}
\includegraphics[width=4cm]{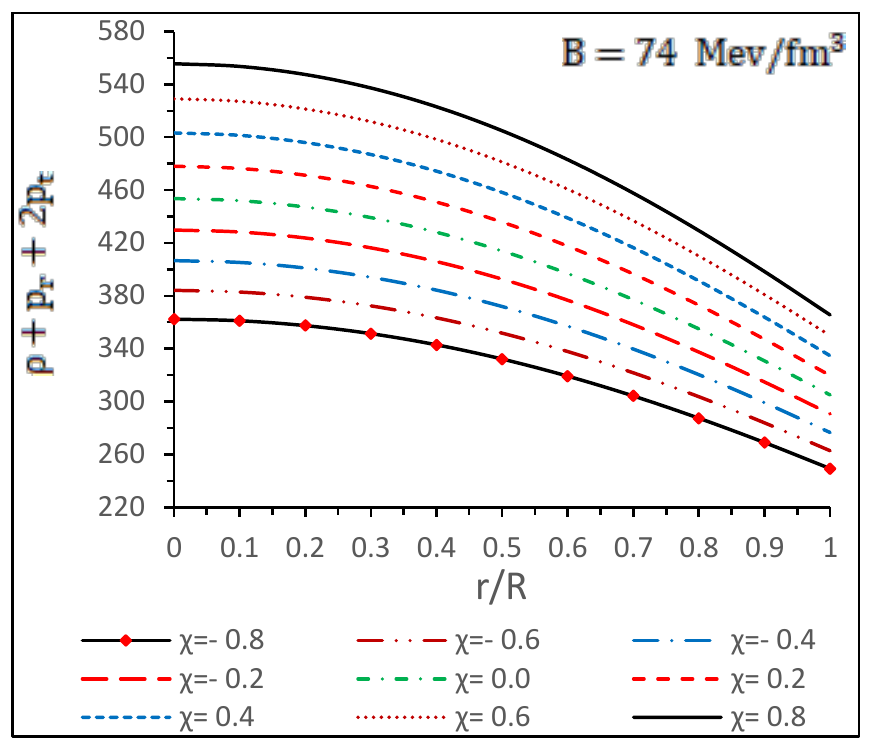}
\includegraphics[width=4cm]{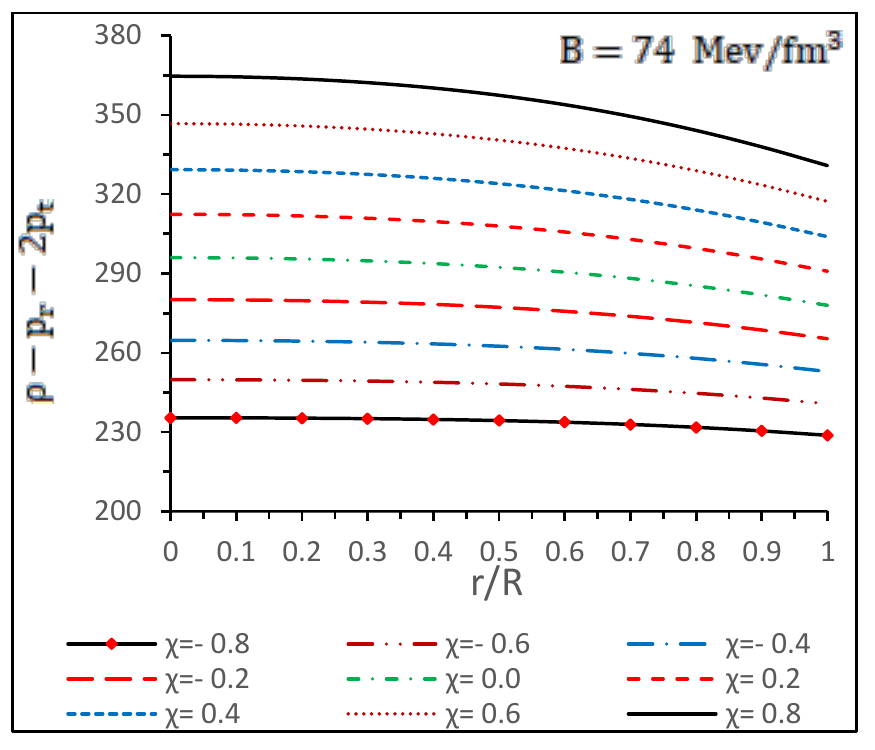}
\caption{Behavior of energy conditions vs. radial   coordinates $r/R$ of SMC X-4 for different values of $\chi$ with bag constant $B=64 Mev/fm^3$ and $B=74 Mev/fm^3$.}\label{Fig4}
\end{center}
\end{figure}
%%%%%%%%%%%%%%%%%%%%%%%%%%%%%%%%%%%%%%%

 The evolution of all these energy conditions against the radial coordinate $r/R$ for the compact stellar structure is well satisfied for our system in the framework of the $f\left(R,\mathcal{T}\right)$ gravity theory at different selected values of $\chi$ and $B$. These are depicted graphically in the Fig.~\ref{Fig4}.

\subsection{Compactness relation and surface gravitational redshift}\label{subsec5.3}
Let us now define the mass function of the compact stellar structure as

\begin{eqnarray}\label{6.1}
 m=\frac{1}{2}\Bigg[\frac{2\, M r^2\,\exp{\left(\frac{M\,(R^2-r^2)}{(2\,M-R)\,R^2}\,\right)}}{ R^2\,(R - 2\, M)+ 2\, M r^2\,\exp{\left(\frac{M\,(R^2-r^2)}{(2\,M-R)\,R^2}\,\right)}}\Bigg] r.
\end{eqnarray}

From this Eq.~(\ref{6.1}), when the radius $r$ is zero, we notice that the mass function is regular at the center of the compact stellar structure. On the other hand, the compactness function of the compact stellar structure can be defined through the equation

\begin{eqnarray}\label{6.2}
 u(r)&=&\frac{m(r)}{r}\nonumber \\
 &=&\frac{1}{2}\Bigg[\frac{2\, M r^2\,\exp{\left(\frac{M\,(R^2-r^2)}{(2\,M-R)\,R^2}\,\right)}}{ R^2\,(R - 2\, M)+ 2\, M r^2\,\exp{\left(\frac{M\,(R^2-r^2)}{(2\,M-R)\,R^2}\,\right)}}\Bigg].
\end{eqnarray}

%%%%%%%%%%%%%%%%%%%%%%%%%%%%%%%%%%
\begin{figure}[h!]
\begin{center}
\includegraphics[width=7.5cm]{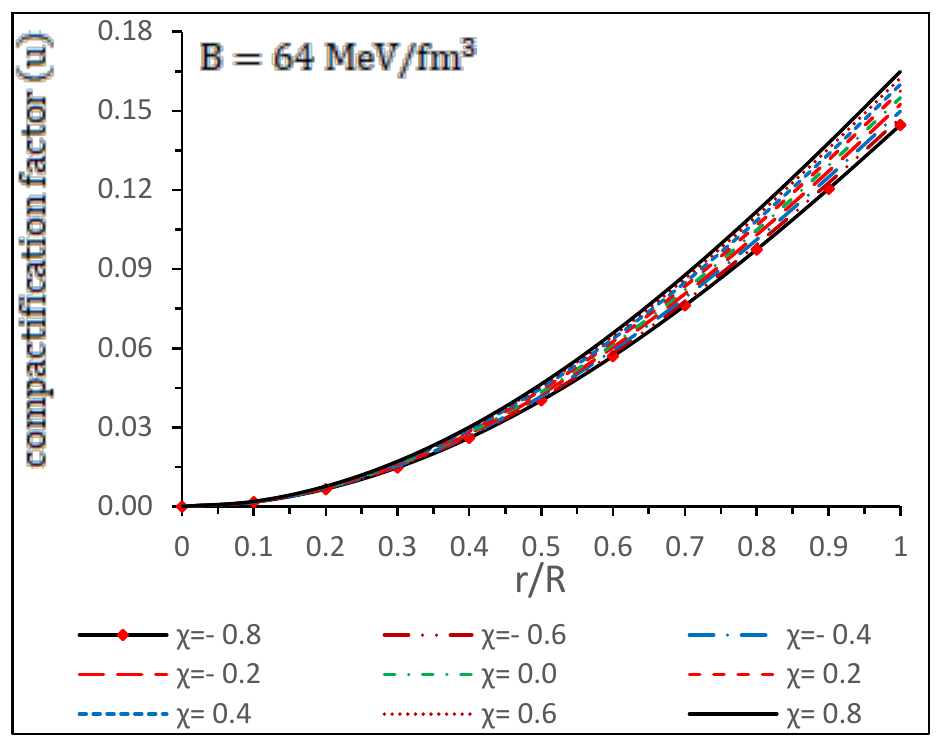}
\includegraphics[width=7.5cm]{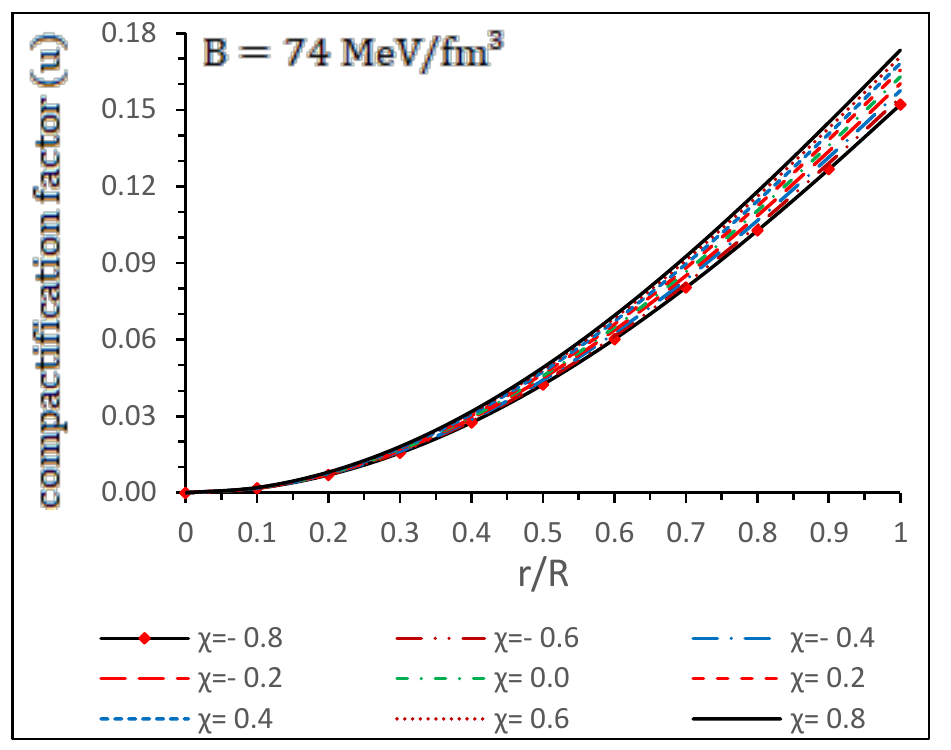}
\caption{Behavior of compactness ($u(r)$) vs. radial coordinates $r/R$ of SMC X-4 for different values of $\chi$ with bag constant $B=64 Mev/fm^3$ and $B=74 Mev/fm^3$.}\label{Fig8}
\end{center}
\end{figure}
%%%%%%%%%%%%%%%%%%%%%%%%%%%%%%%%%%%%%%%

The behavior of the compactness function is illustrated in Fig.~\ref{Fig8}. This figure shows that the compactness function is a monotonic expanding function with respect to the radial coordinates $r/R$. To see most extreme conceivable mass-to-radius ratio, as anticipated by~\citet{Buchdahl1959}, is given by $\frac{M}{R}<\frac{4}{9}$. Fig.~\ref{Fig8} demonstrates that for all the estimations of $\chi$ and $B$, the typical structure of Buchdahl is fulfilled in our present mode.

We have also decided the  gravitational redshift function  $Z$ of a compact stellar structure can be given by the following formula :
\begin{equation}\label{6.3}
Z=-1+\sqrt{1-\frac{2\, M r^2\,\exp{\left(\frac{M\,(R^2-r^2)}{(2\,M-R)\,R^2}\,\right)}}{(2\,M-R)\,R^2}}.
\end{equation}

\begin{figure}[h!]
\begin{center}
\includegraphics[width=7cm]{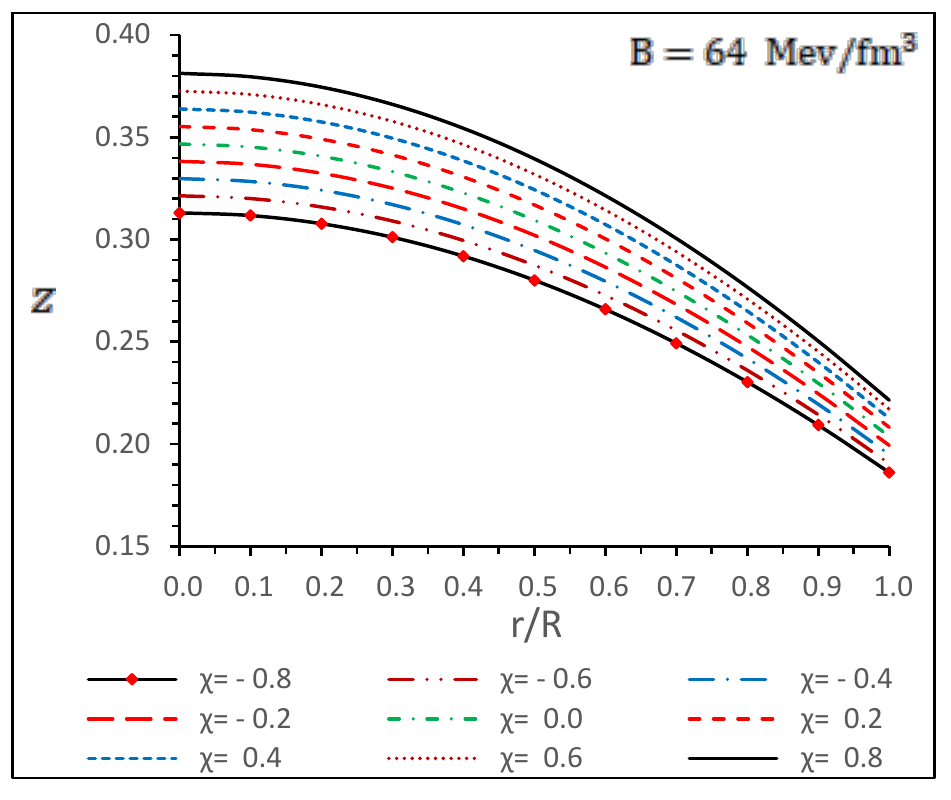}
\includegraphics[width=7cm]{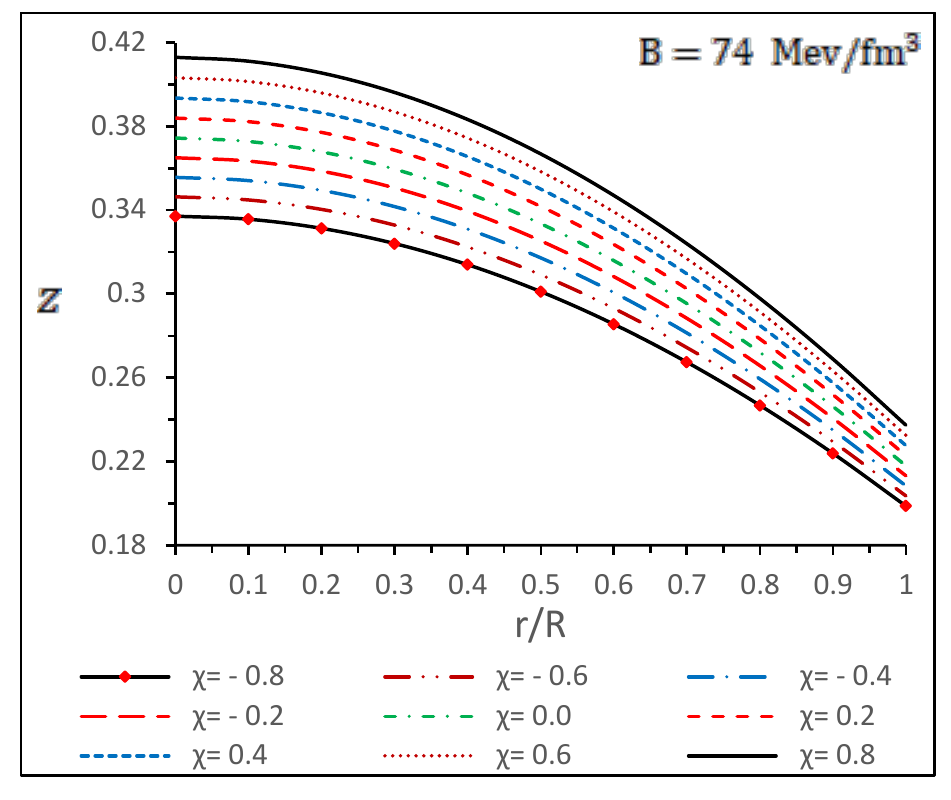}
\caption{Behavior of redshift ($Z$) vs. radial coordinates $r/R$ of SMC X-4 for different values of $\chi$ with bag constant $B=64 Mev/fm^3$ and $B=74 Mev/fm^3$.}\label{Fig9}
\end{center}
\end{figure}

The profile of the gravitational redshift relation $Z$ is represented in Fig.~\ref{Fig9}. This figure exhibits that the gravitational redshift function is monotonically decreasing, as a function to the radius $r/R$, inside the compact stellar structure for all the values of $\chi$ and $B$. Moreover the figure shows that the surface gravitational redshift increases when the value of $\chi$ moves from $-0.8$ to $0.8$.

%%%%%%%%%%%%%%%%%%%%%%%%%%%%%%%%%%
%%%%%%%%%%%%%%%%%%%%%%%%%%%%%%%%%%%%%%%

 \subsection{Dynamical Equilibrium and Stability}\label{5.2}
 In this part, we consider the modified TOV equation in the framework $f\left(R,\mathcal{T}\right)$ theory gravity, causality condition, and adiabatic index in order to perform equilibrium and stability analysis. These approaches would assist us to explore various stability highlights of compact stellar structures are as follows :

\subsubsection{Modified Tolman-Oppenheimer-Volkoff (TOV) equation in the framework $f\left(R,\mathcal{T}\right)$ theory gravity} \label{subsubsec5.2.1}
To simplify the analysis and make the solution more viable, we explore Eq.~(\ref{1.5}) which must meet some general physical requirements of the Einstein field equation for the $f\left(R,\mathcal{T}\right)$ theory gravity. Generally, the field equations also depend, through the conservation equation of the effective stress-energy tensor of our structure as
 \begin{equation}\label{5.2.1.1}
 \nabla^{\mu}{T}^{eff}_{\mu\nu}=0.
 \end{equation}

From this equation, we obtain the modified TOV equation for the spherical anisotropic stellar interior within the framework of the $f\left(R,\mathcal{T}\right)$ gravity theory as follows

\begin{eqnarray}\label{4.9}
 -\frac{\nu^{{\prime}}}{2} \left( \rho+p_r \right)-\frac{dp_r}{dr}+\frac{2}{r}\left({p_t}-{p_r}\right) +{\frac {\chi}{3(8\pi+2\chi)}}\left(3\,\frac{d\rho}{dr}-\frac{dp_r}{dr}-2\,\frac{dp_t}{dr} \right)=0.
\end{eqnarray}

The above modified TOV equation (\ref{4.9}) describes the equilibrium condition for anisotropic fluid in $f(R,T)$ gravity system. The quantity $-\frac{\nu^{{\prime}}}{2} \left( \rho+p_r \right)$ represents the gravitational force ($F_g$), the quantity $-\frac{dp_r}{dr}$ indicates hydrodynamic force ($F_h$), the quantity $\frac{2}{r}\left({p_t}-{p_r}\right)$ denotes anisotropic force ($F_a$) and the quantity ${\frac {\chi}{3(8\pi+2\chi)}}\left(3\,\frac{d\rho}{dr}-\frac{dp_r}{dr}-2\,\frac{dp_t}{dr} \right)$ describes the resultant force (${F_\chi}$) of the coupling between the geometry and the matter. Let us now try to clarify the Eq.~(\ref{4.9}) from an equilibrium perspective. This equation showed that the equilibrium condition must be stable for the compact stellar system if the sum of the four different forces is zero, i.e, the above condition is defined as, ${F_g}+{F_h}+{F_a}+{F_\chi}=0$.

\begin{figure}[h!]
\begin{center}
\includegraphics[width=7.5cm]{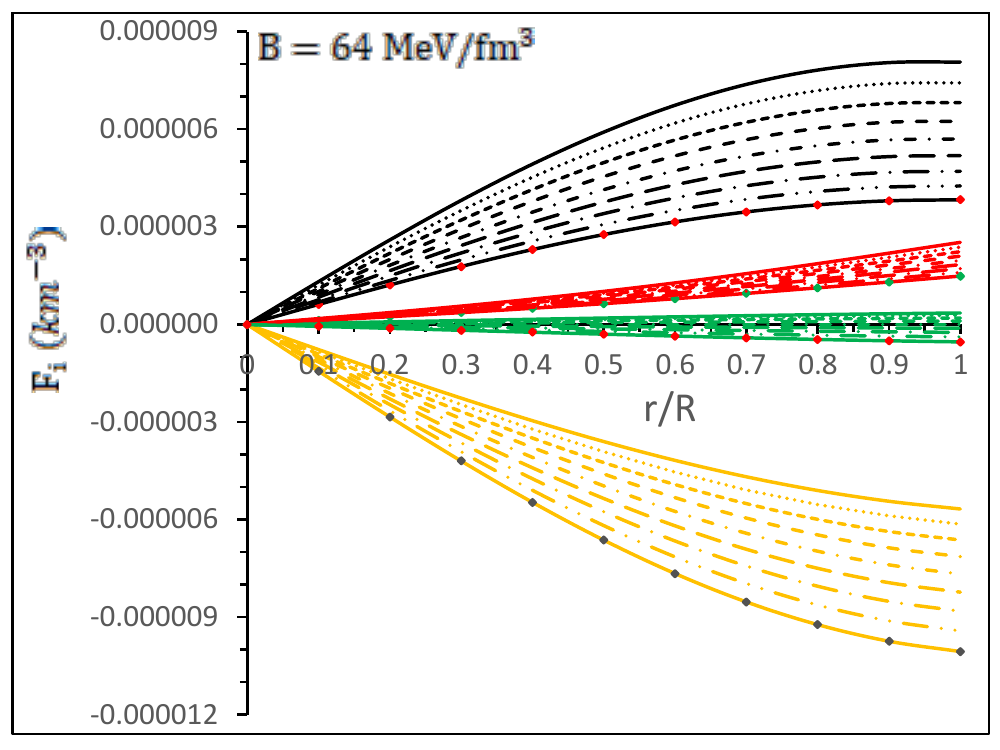}
\includegraphics[width=7.5cm]{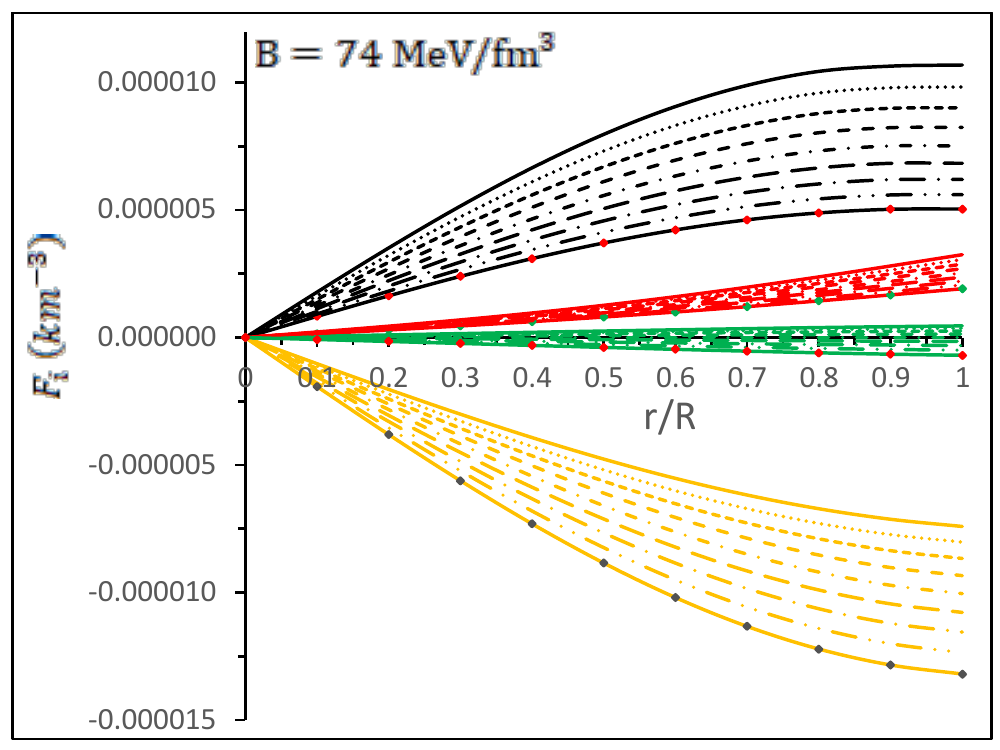}
\caption{Behavior of different forces $F_g$ (yellow lines), $F_h$ (black lines), $F_a$ (red lines) and $F_\chi$ (green lines)  vs. radial coordinates $r/R$ of SMC X-4 for different values of $\chi$ with bag constant $B=64 Mev/fm^3$ and $B=74 Mev/fm^3$. The description of plotted lines are as follows: Solid lines with marker for $\chi=-0.8$, Dashed lines with doubled dot for $\chi=-0.6$, Dashed lines with dot for $\chi=-0.4$, Dashed lines for $\chi=-0.2$, Small dashed lines with dot for $\chi= 0.0$, Small dashed lines for $\chi=0.2$, Dotted lines for $\chi=0.4$, Small dotted lines for $\chi=0.6$, Solid lines for $\chi=0.8$.}\label{Fig5}
\end{center}
\end{figure}

To guarantee the stability of the proposed stellar structure. we have shown in Fig.~\ref{Fig5} that the equilibrium of the forces is reached to all the values of ${\chi}$ and some specific values for ${B}$ which is validated the stability of our model. Consequently, Fig.~\ref{Fig5} indicates that, in the situation of ${\chi}>0$, the resulting impact of hydrodynamic force (${F_h}$), anisotropic force (${F_a}$), and resultant force (${F_\chi}$) of the coupling between matter and geometry compensates the internal attraction due to the gravitational force (${F_g}$). In the situation where ${\chi}<0$, the combined impact of the two anisotropic and hydrodynamic forces is counterbalanced by the sum of the gravitational force and the resultant force of the coupling between the matter and geometry. Hence, for parameter ${\chi}$ to take esteems that are entirely negative we find ${F_\chi}$ acts along the outward direction and behaves like a repulsive force, while for parameter ${\chi}$ takes esteems that are entirely positive the impact of ${F_\chi}$ acts along the inward direction and shows attractive nature. Again we can see from Fig.~\ref{Fig5} that the overall profile of various forces for the compact stellar structure due to chosen parametric values of ${\chi}$, viz., ${\chi=-0.8, -0.6, -0.4, -0.2, 0.0, 0.2, 0.4, 0.6,}$ and ${0.8}$ and the two chosen values of bag constants as $B=64 MeV/fm^3$ and $B=74 MeV/fm^3$. At the point when ${\chi}$ increments from ${\chi=-0.8}$ to ${\chi=0.8}$, we can observe that the values force components $F_g$, $F_h$, $F_a$ and $F_{\chi}$ also increases with   $\chi$.

%%%%%%%%%%%%%%%%%%%%%%%%%%%%%%%%%%
%%%%%%%%%%%%%%%%%%%%%%%%%%%%%%%%%%%%%%%

\subsubsection{The status of the sound speed within the stellar system}\label{subsubsec5.2.2}
For a physically satisfactory model, the squares of the radial and tangential speeds of the sound must be included in the ranges corresponding to the following conditions:~$0<v^2_{r}<1$~and~$0<v^2_{t}<1$~\citep{Abreu2007,Herrera1992}, which can be described as a condition of causality. In addition, for the verification of stability, there is another strategy for the distribution of anisotropic matter, known as the Herrera cracking approach~\citep{Herrera1992}. Now, we use this approach to identify the structure of the anisotropic matter that is likely to be stable or unstable. According to the approach of Herrera~\citep{Herrera1992} and Andr\`{e}asson~\citep{Andreasson2009} for the stability of the distribution of matter, the region for which~$0<\mid v^2_{t} - v^2_{r}\mid <1$ i.e., no cracking is a potentially stable region.  For our stellar system, the radial and tangential speeds of sound are defined respectively as,

\begin{eqnarray}
 v^2_r&=&\frac{dp_r^{eff}}{d\rho^{eff}}\\
 v^2_t&=&\frac{dp_t^{eff}}{d\rho^{eff}}
\end{eqnarray}

%%%%%%%%%%%%%%%%%%%%%%%%%%%%%%%%%%
\begin{figure}[h!]
\begin{center}
\includegraphics[width=5.5cm]{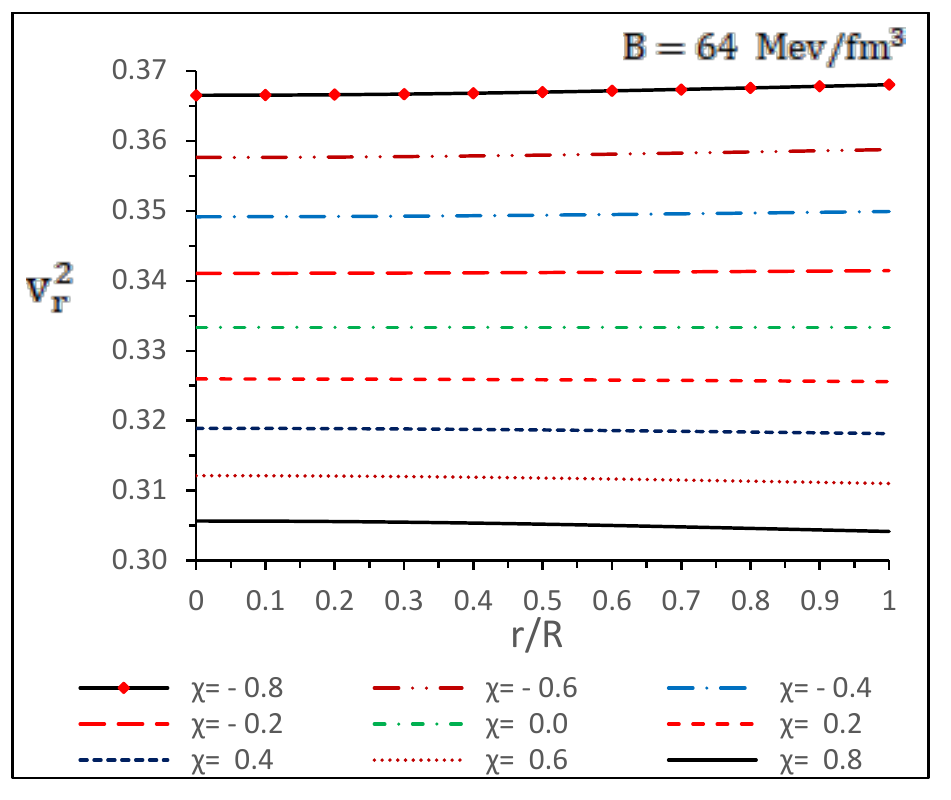}
\includegraphics[width=5.5cm]{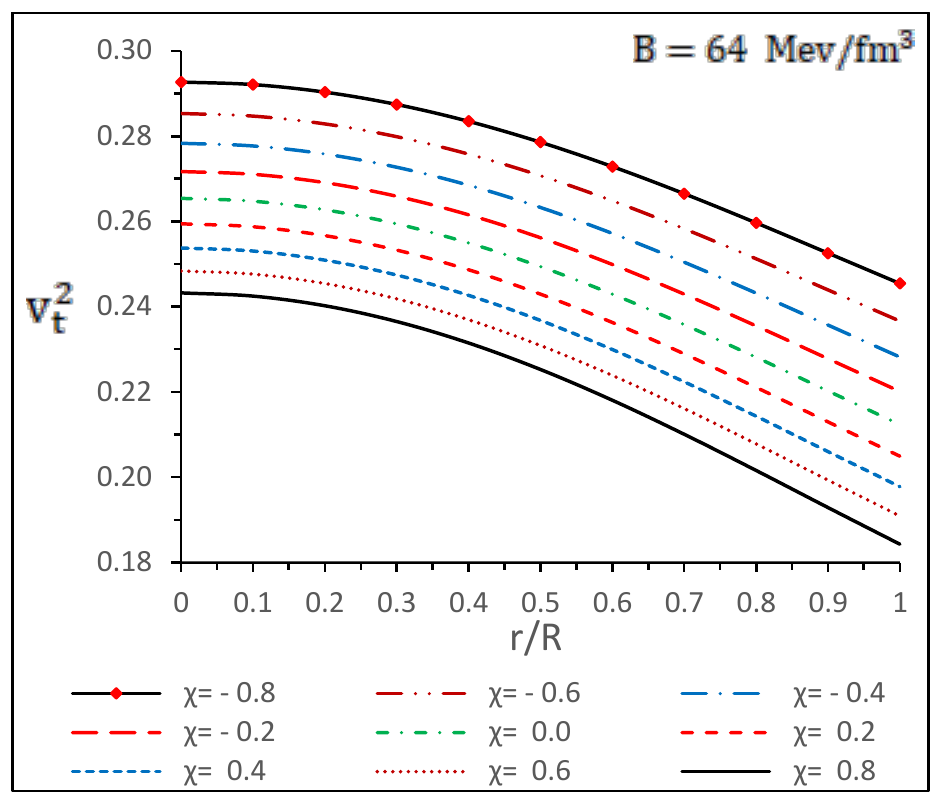}
\includegraphics[width=6.3cm]{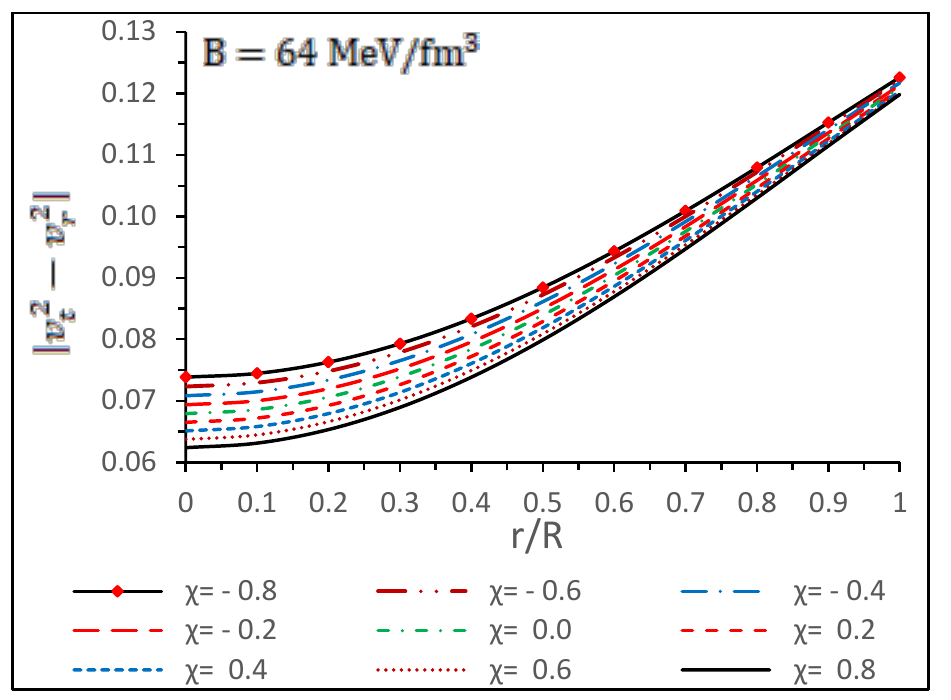}
\includegraphics[width=5.5cm]{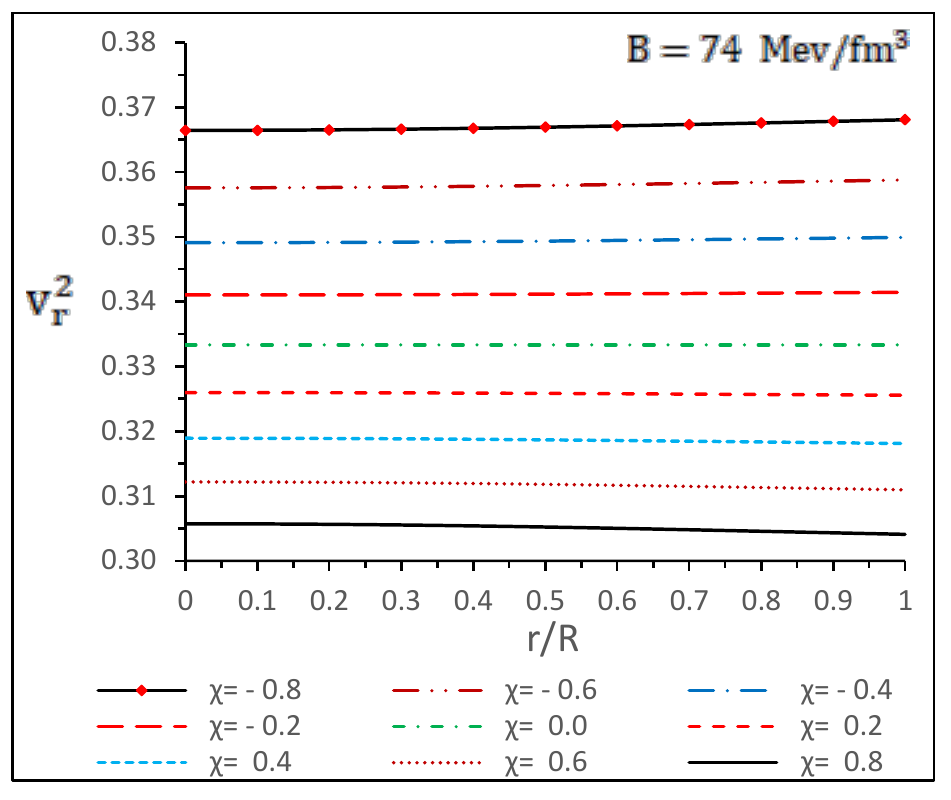}
\includegraphics[width=5.5cm]{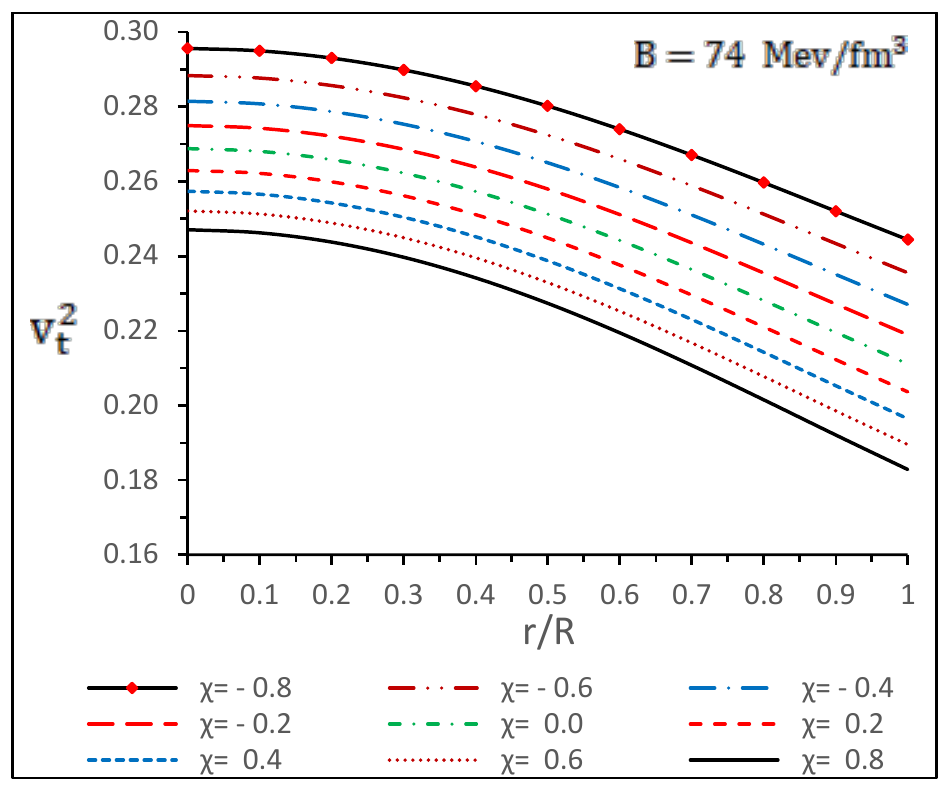}
\includegraphics[width=6.3cm]{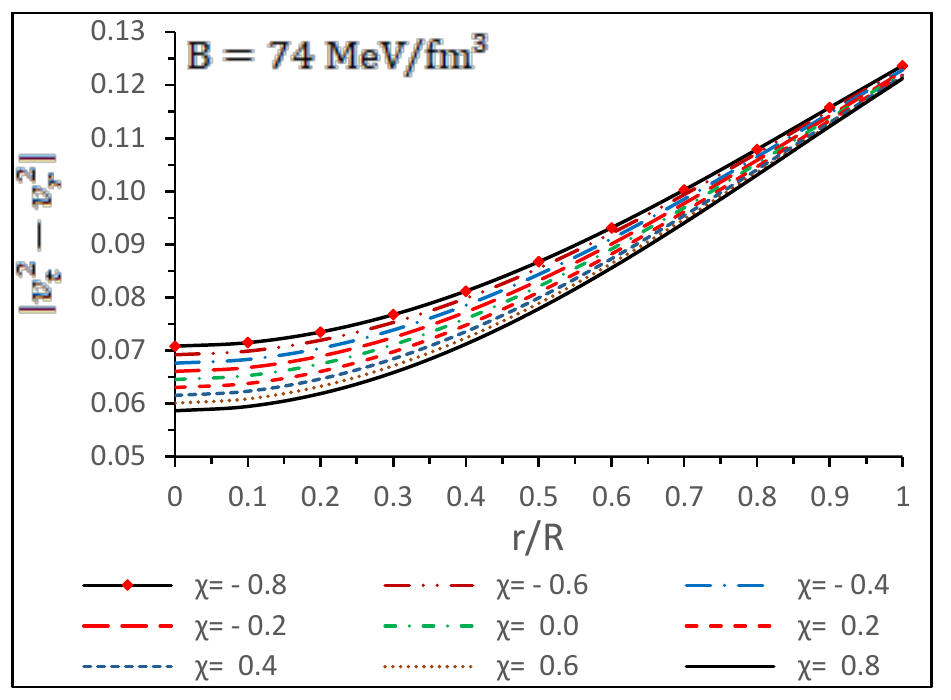}
\caption{Behavior of radial velocity ($v_r^2$), tangential velocity ($v_t^2$) and $|v_t^2-v_r^2|$  vs. radial coordinates $r/R$ of SMC X-4 for different values of $\chi$ with bag constant $B=64 Mev/fm^3$ and $B=74 Mev/fm^3$.}\label{Fig6}
\end{center}
\end{figure}
%%%%%%%%%%%%%%%%%%%%%%%%%%%%%%%%%%%%%%%

The behavior of the square of the radial and tangential sound speeds and their absolute value between them are reported in Fig.~\ref{Fig6} and demonstrates that our framework in $f\left(R,\mathcal{T}\right)$ gravity theory is predictable with both the Herrera cracking approach and the condition of causality, which the validates of our stellar model stability.

\subsubsection{Adiabatic index}\label{subsubsec5.2.3}
The stability of relativistic and non-relativistic spherical objects can be examined by studying the adiabatic index of the stellar system. The investigation of the adiabatic index is essential for the structure of the stellar object with spherical symmetry because it describes the solidity of the equation of state at a given density~\citep{Harrison1965,Haensel2007}. Following Chandrasekhar~\citep{Chandrasekhar1964a,Chandrasekhar1964b} in his pioneering work, several authors~\citep{Hillebrandt1976,Doneva2012,Silva2015,Horvat2010} presented the most elegant method for testing the dynamic stability of stellar objects with symmetry in the face of an infinitesimal radial adiabatic perturbation. According to Heintzmann and Hillebrandt~\citep{HH1975}, an anisotropic compact spherical object model will be stable if $\Gamma>\frac{4}{3}$ everywhere in the interior of the stellar object with spherical symmetry where the adiabatic index $\Gamma$ is defined as
\begin{equation}
\Gamma=\frac{p^{\it eff}_r+{\rho}^{\it eff}}{p^{\it eff}_r}\,\frac{dp^{\it eff}_r}{d{\rho}^{\it eff}}=\frac{p^{\it eff}_r+{\rho}^{\it eff}}{p^{\it eff}_r}\,v^2_{r}. \label{5.2.3.1}
\end{equation}
%%%%%%%%%%%%%%%%%%%%%%%%%%%%%%%%%%
\begin{figure}[h!]
\begin{center}
\includegraphics[width=8.2cm]{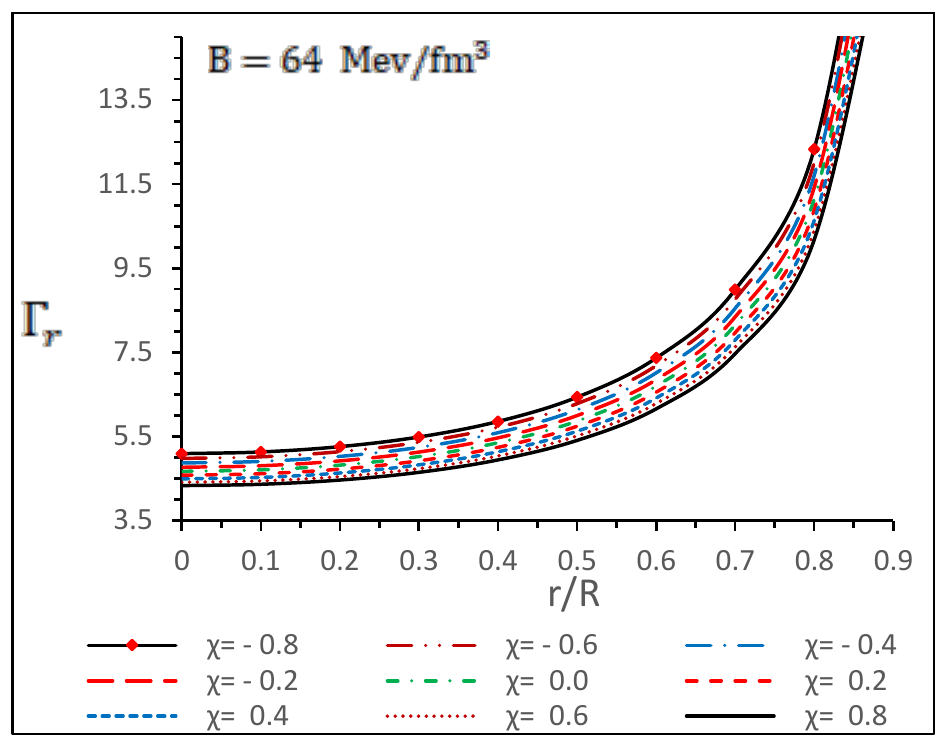}
\includegraphics[width=7.5cm]{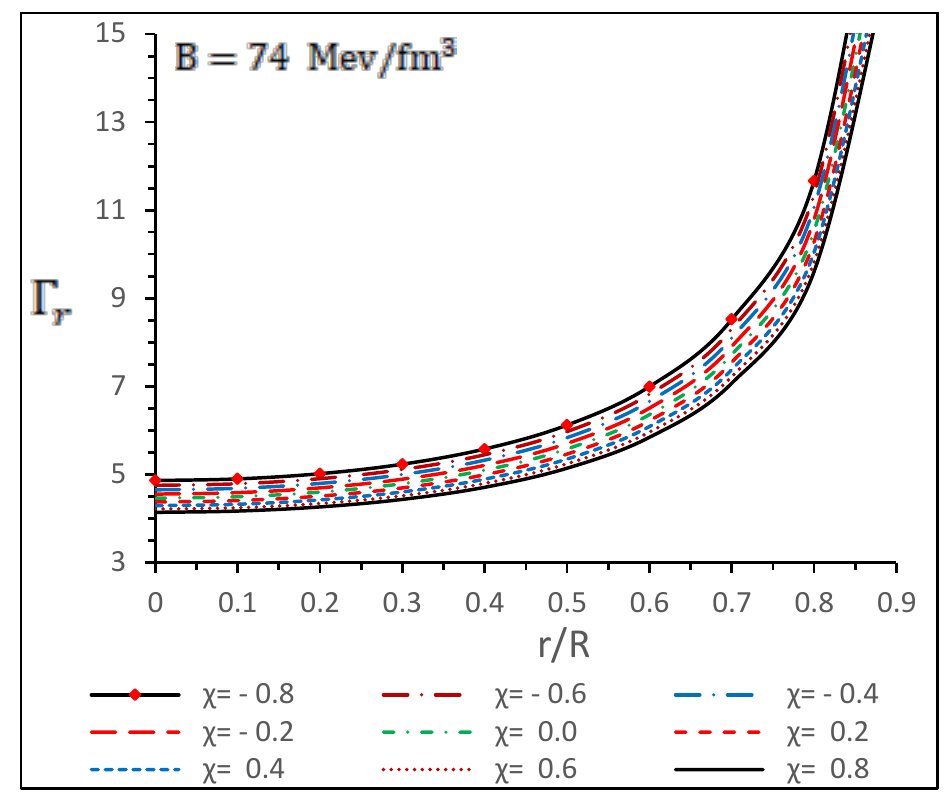}
\caption{Behavior of adiabatic index ($\Gamma$) vs. radial coordinates $r/R$ of SMC X-4 for different values of $\chi$ with bag constant $B=64 Mev/fm^3$ and $B=74 Mev/fm^3$.}\label{Fig7}
\end{center}
\end{figure}

We can see from Fig.~\ref{Fig7} that the variation  of the adiabatic index of our stellar system is greater than $4/3$ everywhere in the stellar system, which completely agrees with the condition of Heintzmann and Hillebrandt~\citep{HH1975}. This characteristic clearly shows that our stellar system is completely stable against infinitesimal radial adiabatic perturbations.
%%%%%%%%%%%%%%%%%%%%%%%%%%%%%%%%%%%%%%%
\begin{table*}
  \centering
    \caption{Physical parameters of the observed strange stars for $\chi=-0.2$ and $B=64 MeV/fm^3$.}\label{Table 1}
        \scalebox{0.95}{
\begin{tabular}{ ccccccccccccccccccccccccccc}
\hline
Strange  & Observed & Predicted  & ${\rho}^{eff}_c$ &  ${\rho}^{eff}_0$ &  $p_c$ & $\frac{2M}{R}$ & $Z_s$ \\
Stars & Mass $(M_{\odot})$ & radius $(Km)$ &  $(gm/{cm}^3)$ & $(gm/{cm}^3)$ & $dyne/{cm}^2$ & & \\
\hline\\
Her X-1 & 0.85$\pm$0.15~\citep{Abubekerov2008} & $9.458^{+0.488}_{-0.554}$  & 4.35287 $\times 10^{15}$  &	5.378 $\times 10^{15}$ & 3.27469 $\times 10^{34}$ & 0.26513 & 0.16653\\

4U 1538-52 & 0.87$\pm$0.07~\citep{Rawls2011}  & $9.527^{+0.231}_{-0.245}$ & 5.44026 $\times 10^{14}$ & 4.35291 $\times 10^{14}$ &3.34044 $\times 10^{34}$ & 0.26944 & 0.169965\\
SAX J1808.4-3658&	0.9 $\pm$0.3~\citep{Elebert2009} & $9.630^{+0.887}_{-1.147}$&	5.47300 $\times 10^{14}$ &4.35297 $\times 10^{14}$ &	3.44083 $\times 10^{34}$ & 0.27596 &0.17522\\
SMC X - 1 &	1.04 $\pm$0.09~\citep{Rawls2011} & $10.067^{+0.258}_{-0.277}$ &	5.62237 $\times 10^{14}$ & 4.35324 $\times10^{14}$ &	3.89876 $\times 10^{34}$ & 0.30483 & 0.19937\\
LMC X - 4 &	1.29 $\pm$0.05~\citep{Rawls2011} & $10.7500^{+0.123}_{-0.128}$ & 5.89511 $\times 10^{14}$ & 4.35372 $\times 10^{14}$ &4.73487 $\times 10^{34}$ &0.3541 & 0.24428\\
EXO 1785-248 &	1.3 $\pm$0.2~\citep{ozel2009}  & $10.775^{+0.472}_{-0.533}$ & 5.90614 $\times 10^{14}$ &	4.35374 $\times 10^{14}$ & 4.76868 $\times 10^{34}$ & 0.35601 & 0.24612\\
Cen X-3   &	1.49 $\pm$0.08~\citep{Rawls2011}  & $11.224^{+0.176}_{-0.183}$ & 6.11889 $\times 10^{14}$ & 4.35410 $\times 10^{14}$ &	5.420907 $\times 10^{34}$ & 0.3916 & 0.28205\\
4U 1820-30 &	1.58 $\pm$0.06~\citep{guver2010b}  & $11.422^{+0.126}_{-0.131}$ &	6.22241 $\times 10^{14}$ &4.35427 $\times 10^{14}$ &	5.73828 $\times 10^{34}$ &0.40814 & 0.29985\\
PSR J1903+327 & 1.667 $\pm$0.021~\citep{Freire2011} &	$11.604^{+0.043}_{-0.0436}$  & 6.32348 $\times 10^{14}$ & 4.35443 $\times 10^{14}$ &	6.04812 $\times 10^{34}$ & 0.42384 & 0.31743\\

4U 1608-52 & 1.74$\pm$0.14~\citep{guver2010a}  & $11.751^{+0.268}_{-0.286}$  &	6.409589 $\times 10^{14}$  &	4.35457 $\times 10^{14} $ &	6.312103 $\times 10^{35}$ & 0.436867 & 0.33258\\

Vela X-1  & 1.77$\pm$0.08~\citep{Rawls2011} &   $11.81^{+0.153}_{-0.159}$  &	6.44535 $\times 10^{14}$ & 4.35462 $\times 10^{14}$ &	6.42172 $\times 10^{34}$ & 0.442188 & 0.33892\\

PSR J1416-2230 & 1.97$\pm$0.04~\citep{Demorest2010} & $12.182^{+0.071}_{-0.071}$  & 6.68815 $\times 10^{14}$ &	$4.35499\times 10^{14}$ &	7.16609 $\times 10^{34}$ & 0.47702 & 0.38279\\\\
\hline
\end{tabular}}
  \end{table*}

\begin{table*}[h!]
  \centering
    \caption{Derived values of constants due to the different strange star candidates for $\chi=-0.2$ and $B=64 MeV/fm^3$.}\label{Table 2}
       \scalebox{1.}{
\begin{tabular}{ ccccccccccccccccccccccccccc}
\hline
Strange Stars   & $A$ ~~& ~~$B$ & ~~$C$ & ~~$D$   \\ \hline
Her X-1 &  $0.001008315286039$ & $ 3.373920859330568 \times10^{2}$ & $0.613570016070468$ & $3.339760727168075$ \\

4U~1538-52 & $0.001015873682637$ & $3.368573174774726\times {10}^{2}$ & $0.607528304469978$ & $23.326382555539476$ \\
SAX J1808.4-3658 & $0.001027490168257$ & $3.360482305834711\times {10}^{2}$ & $0.598406266725289$ & $3.305943321601397$ \\
SMC~X-1 & $0.001081678451402$  &  $3.324667783235598 \times {10}^{2}$ & $0.558312030374487$ & $3.212501963028448$ \\
LMC~X-4 & $0.0011860165817161$ &  $3.263521330068053 \times {10}^{2}$ & $0.491037682702438$ & $3.040969194751356$ \\
EXO 1785-248 & $0.001190391892880$  & $3.261154961490233 \times {10}^{2}$ & $0.488466814342001$ & $3.034006051856014$ \\
Cen X-3 & $0.001277329046604$  &  $ 3.216990833610159 \times {10}^{2}$ & $0.440978988700054$ & $2.899282214844905$ \\
4U 1820-30 & $0.001321462557105$ &  $3.196464075782649 \times {10}^{2}$ & $0.419246959742855$ & $2.833443744919449$ \\
PSR J1903+327 & $ 0.001365773970524$ &  $3.176992097631392 \times {10}^{2}$ & $0.398847541082252$ & $2.768993062638951$ \\
4U 1608-52 & $0.001404522601536$   & $3.160824246435486 \times {10}^{2}$ & $0.382078429900167$ & $2.713948382888694$ \\
Vela X-1  & $0.001420890658661$   & $3.154220804145845 \times {10}^{2}$ & $0.375275357251443$ & $2.691056280652941$ &\\
PSR J1416-2230 & $0.001536581471130$ & $3.110999490848716 \times {10}^{2}$ & $0.331451335091942
$  &	$2.535101121164858$ \\
\hline
\end{tabular}  }
  \end{table*}

%%%%%%%%%%%%%%%%%%%%%%%%%%%%%%%%%%%%%%%%%%%%%%%%%%%%%%%%%%%%%%%%%%%%%%%%%%%%%%%%%%%%%%%%%%%%%%%%

\section{Discussion and conclusion}\label{sec6}

In this paper, we have studied a new class of generalized anisotropic solutions for compact stellar structures with spherically symmetric, known as strange spherical objects. In our study, to obtain general solutions to modified Einstein's field equations in the context of $f\left(R,\mathcal{T}\right)$ gravity, we emphasize the technique of embedding class as a ground-breaking tool. To be specific, our survey proceeds in the framework of class 1 of the embeddings, by techniques for embedding within 4-dimensional space-time into a 5-dimensional flat Euclidean space, so as to obtain a solution to the modified Einstein field equations in the existence of framework of $f\left(R,\mathcal{T}\right)$ gravity theory. In light of the improved linear form of the arbitrary function of the $f\left(R,\mathcal{T}\right)$ theory gravity writen in the spesific form $f\left(R,\mathcal{T}\right)=R+2\chi\mathcal{T}$, to incorporate the interaction between the  matter and geometrical terms, which has been provided by Harko et al.~\citep{harko2011}. In Eq.~(\ref{1.5}) we provide an anisotropic extension of field equation due to the modified action of Einstein-Hilbert $f\left(R,\mathcal{T}\right)$ gravity, which makes it clear that our stellar structure is not just a strange quark matter, but it presents another type of unknown as a coupling effect of curvature-matter. In the same spirit, Chakraborty~\citep{SC2013} has examined the nature and origin of this new sort of matter which is at the birthplace of a particular interaction between the geometry and matter. Currently considering the effective stress-energy tensor of the distribution of matter like $T^{eff}_{\mu\nu}$ in Eq.~(\ref{1.5a}) that outcomes in $\nabla^{\mu}T^{eff}_{\mu\nu}=0$, it can be settled the non-conservation of the stress-energy tensor presented in Eq.~(\ref{1.6}).To progress toward the solutions of the gravitational field Eqs.~(\ref{2.3})-(\ref{2.2}), we have proposed the simplified phenomenological MIT bag model for strange quark matter. It is basically described by the equation of state indicated as $p_r=\frac{1}{3}\left(\rho-4B\right)$, with the use of the embedding class 1 methods~(\ref{3.2}).Consistent with the idea similar to Lake~\citep{Lake2003}, we chose the gravitational metric potential in Eq.~(\ref{3.1}) that meets all the requirements, then we use embedding class one condition~(\ref{3.3}), we get the metric function given in Eq.~(\ref{3.4}).

All through the investigation, we have been considering SMC~X-1 of the observed mass	1.04 $\pm$0.09~$M_{\odot}$~\citep{Rawls2011} as the delegate of the strange quark stars. Now, we suppose that the arbitrary estimations of ${\chi}$ were ${\chi=-0.8, -0.6, -0.4, -0.2, 0.0, 0.2, 0.4, 0.6,}$ and ${0.8}$, and that the two estimations of the bag constant $B$ as $64 MeV/fm^3$ and $74 MeV/fm^3$, which corresponds to the acknowledged estimations of $B$~\citep{Burgio2002, Kalam2013, Rahaman2014}, in order to determine the unknown values of the different arbitrary constants and parameters, namely, $A$, $B$, $C$ and $D$, etc., and the radius $R$ of the complete structure of the stellar models $f\left(R,\mathcal{T}\right)$ gravity.

The main physical highlights of the curent solution can be utilized to investigate a new class of generalized anisotropic solutions for the strange spherical objects as follows:

\begin{enumerate}
    \item Firstly, it can be seen in Fig.~\ref{Fig1} that the metric potentials, namely, ${\rm e}^{\nu}$ and ${\rm e}^{\lambda}$, with respect to the radial coordinate $r/R$, exhibit a monotonically increasing behavior of the two metric potentials of the original finite values at the boundary values at the surface. The profile of physical quantities of effective energy density, effective radial pressure, and tangential pressure, viz, ${\rho}^{eff}$, $p_r^{eff}$ and $p_t^{eff}$, respectively, with respect to the radial coordinate $r/R$, are represented in  Fig.~\ref{Fig2}, which shows that all the physical quantities have maximum values at the origin decreasing progressively to reach the minimum values at the surface as well as to approve the physical availability of the obtained solutions. From the Figs.~\ref{Fig1}~and~\ref{Fig2} we confirm again that our stellar structure is totally devoid of physical or geometrical singularities for some different parametric values of $\chi$ and $B$.
    \item The plot corresponding to the behavior of effective anisotropy $\Delta_{eff}$ against radial coordinates $r/R$ is depicted in Fig.~\ref{Fig3}. We see that the effective anisotropy of the system increases as the radius increases. For instance, $\Delta_{eff}$ is zero in the center of the system and increases to reach the maximum value at the surface of the system. On the other hand, in the existence of the framework of $f\left(R,\mathcal{T}\right)$ gravity, it is interesting to note that our effective anisotropy follows the same profile as anticipated by Deb et al.~\citep{Deb2017} in the general relativity case.
    \item Fig.~\ref{Fig.6} illustrates the behavior of the total mass $M$, normalized in solar mass units $M_{\odot}$ with the total radial coordinate $R$ of our model for strange spherical object candidates. We have shown, for various estimates of $\chi$ and $B$, where $B$ is the bag constant, that the mass-radius function for strange compact objects in $f\left(R,\mathcal{T}\right)$ gravity theory achieved typical profile as in general relativity. Moreover, we find that for the parameter values of $\chi$ moved from $-0.8$ to $0.8$ gradually, the estimates of the radius decreases progressively, which indicates that our anisotropic stellar structure becomes more massive and transforms into more dense compact stars.
    \item Further, we examined the energy conditions, the mass-radius relationship, and the stellar structure stability, etc., in order to test the physical validity of the solutions obtained. In Fig.~\ref{Fig4} we have indicated the behavior of all energy conditions with respect to the radial coordinate $r/R$ for the stellar system, which shows that our compact stellar structure is well satisfied for the system in the context of the gravity $f\left(R,\mathcal{T}\right)$ at various choose values of $\chi$ and $B$. The numerical results corresponding to the compactification factor and gravitational redshift are given in Table \ref{Table 3} and \ref{Table 4} while the behaviors of the compactness relation and the redshift function are shown by Figs.~\ref{Fig8}~and~\ref{Fig9} respectively for $B=64 MeV/fm^3$ and $B=74 MeV/fm^3$.
    \item We have shown in Fig.~\ref{Fig5} that the equilibrium of the forces is reached to all the values of $\chi$ and some specific values for $B$ which confirms that our stellar model is stable with respect to the equilibrium of forces. Consequently,  Fig.~\ref{Fig5} indicates that, for $\chi<0$, we find $F_{\chi}$ acts along the outward direction and behaves like a repulsive force, while for $\chi>0$, the impact of $F_{\chi}$ acts along the inward direction and shows attractive nature. The current anisotropic stellar system is furthermore predictable with the Herrera~\citep{Herrera1992} and Andr\`{e}asson~\citep{Andreasson2009} approach and causality condition for the stability of the distribution of matter as a profile of the difference in squared of velocities of sound,~$\mid v^2_{t} - v^2_{r}\mid$ whith respect to the radial coordinate $r/R$ satisfies the inequality~$0<\mid v^2_{t} - v^2_{r}\mid<1$ which manifests itself in Fig.~\ref{Fig6}.
    \item Further, in Fig.~\ref{Fig7} we have displayed the behavior of the adiabatic index~$\Gamma$ with respect to the infinitesimal radial adiabatic perturbation which confirms that when~$\Gamma> 4/3$ our stellar structure is stable in all interior points of the stellar object with spherical symmetry.
    \item On the one hand in Table~\ref{Table 1}, we predicted several values of the physical parameters of the strange stars observed for $\chi=-0.2$ and $B=64 MeV/fm^3$, namely, the surface redshift $Z_s$, the mass ratio $2M/R$, the central pressure $p_c$, the surface density ${\rho}_0^{eff}$, the central density ${\rho}_c^{eff}$ and the radius $R$. On the other hand, we confirm that the selected stellar structures are suitable candidates for the ultra-dense hypothetical strange stars through the results given in Table~\ref{Table 1}, such as the values of high redshift and their surface densities are greater than the normal nuclear density as indicated by Ruderman~\citep{Ruderman1972}, Glendenning~\citep{Glendenning1997}, and Herzog and Ropke~\citep{Herzog2011}. Next, in Table~\ref{Table 2}, we derived the values of the constants due to the different strange star candidates for the same values of $\chi$ and $B$ shown in Table~\ref{Table 1}, viz., $A$, $B$, $C$ and $D$. Further, we presented the physical properties of the SMC X-1 due to different values of $\chi$ for the two values of Bag constants $B$ such as $B=64 MeV/fm^3$ and $B=74 MeV/fm^3$, are shown respectively in Tables~\ref{Table 3}~and~\ref{Table 4}, in order to obtain a clear motivation for the comparative discussion. And also, in Table~\ref{Table 5} we have predicted the physical properties of the SMC X-1 due to different values of $B$ viz., $B=56, 86$~and~$96$ for $\chi=0.2$, which shows the predicted values of $Z_s$, $2M/R$, $p_c$, ${\rho}_0^{eff}$, ${\rho}_c^{eff}$ and $R$ corresponds to these three chosen values of $B$. Furthermore, Tables~\ref{Table 3},~\ref{Table 4}~and~\ref{Table 5} they clearly indicate that the physical quantities $Z_s$, $2M/R$, $p_c$, ${\rho}_0^{eff}$, and ${\rho}_c^{eff}$ are always increasing when $\chi$ and $B$ increase gradually, accordingly, the radius $R$ decreases, which obviously shows that the anisotropic stellar structures enlarge and become more dense compact objects.
\end{enumerate}

We have exhibited a comparative and intriguing outcome with that predicted by Astashenok and his collaborators~\citep{Astashenoka2015} in the framework of the $f\left(R,\mathcal{T}\right)$ gravity theory for a specific form given by Eq.~(\ref{frt}), to take into account the simplest coupling between matter and geometry. Further investigation by Astashenok et al.~\citep{Astashenoka2015} gives a nonperturbative model of strange spherical objects in $f(R)=R+2\,\alpha\,R^2$ gravity theory, where $\alpha$ is a constant. They have shown that the mass of candidate strange spherical objects increases when the value of the parameter $\alpha$ increases progressively, which in our investigation is a fascinating case because of the diminishing estimations of the parameter $\chi$. It is fascinating to take note of that the additional gravitational mass was emerging on account of $f(R)$ gravity theory~\citep{Astashenoka2015} because of the additional geometrical term $2\,\alpha\,R^2$, while in our investigation the equivalent is acquired because of the additional matter term $2\,\chi\,\mathcal{T}$. Astashenoka, Capozziello and Odintsov~\citep{Astashenoka2015} found that in the context of the $f(R)$ gravity, the values of physical quantity $Z_s$ decrease accordingly with the increase of the values of $\alpha$, whereas in our case we find within the framework of the theory of the particular $f\left(R,\mathcal{T}\right)$ gravity, when the values of $\chi$ increase, the values of physical quantity $Z_s$ increase progressively. Therefore, it is easy to distinguish the effects of the gravity $f\left(R,\mathcal{T}\right)$ and $f(R)$ by studying the physical quantity known as the surface redshift of compact stellar systems.

The validity of Einstein's theory, which has been successfully tested mainly under the feeble gravity regime by means of research center investigations and solar system tests, is highly confirmed and still faces a strict constraint in the region near the black hole, the ultra-dense compact objects and the expanding universe through the strong gravity regime. However, observations of several peculiar, over- and under-luminous type Ia supernovae e.g. SN 2003fg, SN 2006gz, SN 2007if and SN 2009dc~\citep{Howell2006,Scalzo2010} demonstrates a gigantic Ni-mass and affirms significantly super-Chandrasekhar limiting mass white dwarfs as progenitors for peculiar over-luminous type Ia supernovae. Lately, Linares et al.~\citep{Linares2018} have found an exceedingly gigantic pulser of mass ${2.27}_{-0.15}^{+0.17}$~$M_{\odot}$ as they would see it of compact binaries PSR J2215 + 5135. Obviously, these perceptions are not simply to examine Chandrasekhar's standard limit for compact stellar structures but also to invoke the need to modify general relativity in the regime of strong gravity. Curiously, our investigation uncovers the maximum mass limits exceed their standard estimates in general relativity for the chosen parametric estimates of $\chi$ due to the impact of the $f\left(R,\mathcal{T}\right)$ theory of gravity. Therefore, one can also explain the massive stellar structures observed, namely, the massive pulsars, Super Chandrasekhar stars and magnetars, etc., through the stellar structures in the framework of the $f\left(R,\mathcal{T}\right)$ theory of gravity, on the other hand, general relativity can hardly be explained so far. However, in the limit $\chi=0$, we can recover Einstein's standard gravity solutions.

Finally, our anisotropic generalized model on ultra-dense compact stellar stars consist of quark matter $(u)$, $(d)$ and $(s)$ are called strange stars survives from all the critical physical tests conducted in the current study and successfully explains all the  effects in the framework of $f\left(R,\mathcal{T}\right)$ gravity theory.  \\\\

\textbf{Acknowledgment:} S. K. Maurya acknowledge continuous support
and encouragement from the administration of University of Nizwa. F. Tello-Ortiz thanks the financial support of the project ANT-1756 at the Universidad de Antofagasta, Chile.

\end{document}